\documentclass[reprint,aps,prb,superscriptaddress,amsmath,amssymb,floatfix,nodate]{revtex4-2}

\usepackage[T1]{fontenc}
\usepackage{graphicx}
\usepackage{dcolumn}
\usepackage{bm}
\usepackage{tikz}
\usepackage[colorlinks=true,linkcolor=blue!60!black,citecolor=blue!60!black,urlcolor=blue!60!black]{hyperref}

\newcommand{\msub}{m_{\mathrm{sub}}}
\newcommand{\avtraj}[1]{\langle #1\rangle_{\mathrm{traj}}}
\newcommand{\Sent}{S_{\mathrm{ent}}}

\newcommand{\ket}[1]{|#1\rangle}
\newcommand{\bra}[1]{\langle #1|}

\begin{document}

\title{Discrete time crystals stabilized by bounded measurement noise}

\author{Pallabi Chatterjee}
\email{ph22d001@iittp.ac.in}
\affiliation{Indian Institute of Technology Tirupati, Tirupati 517619, India}

\author{Tanay Nag}
\email{tanay.nag@hyderabad.bits-pilani.ac.in}
\affiliation{Department of Physics, BITS Pilani-Hyderabad Campus, Telangana 500078, India}

\author{Ranjan Modak}
\email{ranjan@iittp.ac.in}
\affiliation{Indian Institute of Technology Tirupati, Tirupati 517619, India}


\begin{abstract}
Periodically driven clean interacting systems generically heat up, which destroys discrete time-crystalline order. Here we ask whether continuous weak measurement can stabilize such order. We study a long-range interacting spin chain whose local occupation is continuously monitored. Measurements generally induce a transition from volume-law to area-law entanglement. Yet low entanglement alone is not enough: along typical quantum trajectories, the system still reaches a featureless state with no subharmonic response. In contrast, the most-probable trajectory, obtained by suppressing measurement fluctuations, shows robust period doubling. To connect these limits, we introduce a bounded-noise protocol that restricts measurement fluctuations to a finite window. Narrow windows preserve period doubling. Intermediate windows produce coexisting time-crystalline and non-time-crystalline trajectories. Over time, trajectories leave the time-crystalline group one by one, so the order is long lived but finite. Wide windows destroy the order. Our results establish bounded measurement noise, discarding the large-deviation trajectories, as a route to time-crystalline order in clean systems. They also show that the measurement-induced area-law phase is necessary, but not sufficient, for this order.
\end{abstract}

\maketitle

\section{Introduction}
\label{sec:intro}

\begin{figure*}[t]
  \centering
  \includegraphics[width=\textwidth]{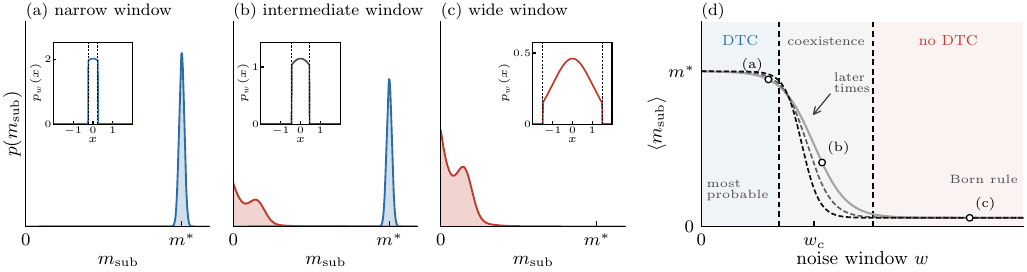}
  \caption{Schematic summary of the main results. (a)--(c) Distribution $p(\msub)$ of the late-time subharmonic response $m^{(r)}_{\mathrm{sub}}$ of individual trajectories for a narrow, an intermediate, and a wide postselection window $w$. Insets: truncated noise distribution $p_w(x)$ of $x=dW/\sqrt{\gamma\,dt}$ [Eq.~\eqref{eq:Pw}], as in the insets of Fig.~\ref{fig:hist}; dashed lines mark the window $|x|<w$. For narrow windows, all trajectories are time crystalline and sit at the most-probable-trajectory value $m^{*}$; at intermediate $w$, a DTC class at $m^{*}$ coexists with a non-DTC class at $\msub\approx0$; for wide windows, only the non-DTC class remains. (d) Trajectory-averaged response $\avtraj{\msub}$ versus $w$: it stays at $m^{*}$ up to $w\approx w_c$ and drops to nearly zero beyond, recovering the featureless Born-rule ($w\to\infty$) limit. Vertical dashed lines separate the DTC, coexistence, and no-DTC regimes (cf.\ Fig.~\ref{fig:nend}). In the coexistence region, $\avtraj{\msub}$ decreases with time (dashed curves) as trajectories switch one by one from the DTC to the non-DTC class. Open circles mark the windows of panels (a)--(c).}
  \label{fig:schematic}
\end{figure*}

Discrete time crystals (DTCs) are nonequilibrium phases in periodically driven isolated systems in the presence of interaction and disorder, manifested through a persistent subharmonic response at an integer fraction of the drive frequency~\cite{DTC_th_1,DTC_th_2,DTC_th_3,DTC_th_4,DTC_th_5,DTC_th_6,DTC_th_7,DTC_th_8,DTC_th_9,DTC_th_10,DTC_th_11,khemani2021comment,Zaletel2023, agrawal2026quantum}. In a conventional period-doubling response, an observable oscillates at a frequency $f_d/2$ when the system is driven at $f_d$. For a genuine DTC, however, the subharmonic response is rigid against small variations of the driving parameters and persists for times much longer than the microscopic timescales. This rigidity distinguishes time-crystalline order from ordinary nonlinear period-doubling phenomena and has motivated extensive studies of its microscopic mechanisms and experimental realizations~\cite{exp_nature_1,exp_nature_2,exp_3,exp_4,exp_5,exp_th_1,exp_th_2,exp_th_3,exp_th_4,exp_th_5,exp_th_6,exp_th_7,ext_th_8,exp_th_9,exp_th_10,exp_th_11,exp_th_12}. Note that the DTC phases due to their spontaneous breaking of discrete time-translation symmetry, are characteristically different from 
other Floquet phases observed in topological systems \cite{oka2019floquet,mondal2026floquet,ghosh2024generation,prnx-47mk}, 
energy pumping \cite{PhysRevX.7.041008,prg5-3srn}, quantum thermodynamics \cite{mori2023floquet,PhysRevB.109.184309,PhysRevE.106.024110,cuetara2015stochastic}.

One of the earliest and most extensively studied mechanisms for stabilizing DTC order is many-body localization (MBL). In MBL systems, strong disorder prevents an interacting many-body system from thermalizing: the system retains a local memory of its initial state, and the eigenstate thermalization hypothesis is violated~\cite{Basko2006,Pal2010,Nandkishore2015,Abanin2019,PhysRevE.101.052108,PhysRevLett.115.230401}. This nonergodicity is reflected in an emergent set of quasilocal integrals of motion, area-law entangled eigenstates, and a slow logarithmic growth of entanglement after a quench~\cite{Abanin2019}. Remarkably, MBL can survive periodic driving, so a driven system need not heat up to infinite temperature~\cite{Ponte2015,Lazarides2015}. In periodically driven interacting disordered systems, localization therefore suppresses thermalization and prevents the system from absorbing energy indefinitely from the drive, allowing long-time memory and time-translation symmetry breaking to persist~\cite{DTC_th_8,DTC_th_11,exp_th_5}. The stability of the resulting phase is therefore intimately related to the nonergodic structure of the Floquet eigenstates, leading to phase-locked oscillations in the order parameter despite imperfections, weak noise, or small changes in drive
parameters. The DTC behavior, however, is not restricted to the MBL regime. Prethermal DTCs can arise in closed systems when a large separation of timescales suppresses heating for parametrically long times~\cite{Else2017,Machado2020}, while dissipative systems provide yet another route in which coupling to an environment can stabilize a subharmonic response~\cite{open_DTC_4}.

The role of interactions and disorder in stabilizing time-crystalline order becomes particularly interesting in systems with long-range interactions. Systems with uniform long-range interactions are known to host prethermal DTC phases in specific parameter regimes, with a strong dependence on the initial state~\cite{exp_5,Machado2020,doi:10.1126/science.abg8102,prethermal_2}. In periodically driven quantum Sherrington--Kirkpatrick models, robust DTC order emerges for random all-to-all interactions and is closely tied to the nonergodic character of the Floquet eigenstates, although nonergodicity alone does not guarantee time-crystalline order~\cite{Bothra2025}. These results highlight the role of eigenstate nonergodicity in stabilizing DTC order in closed quantum systems.

While most studies of DTCs have considered isolated quantum systems, the interplay between time-crystalline order and environmental coupling has also received considerable attention \cite{open_DTC_4}. Although dissipation generally destroys the coherent dynamics of the underlying DTC order~\cite{open_DTC_destroy_1,open_DTC_destroy_2}, suitably engineered dissipative processes can instead stabilize persistent subharmonic oscillations by balancing the energy absorbed from the periodic drive. Such dissipative time-crystalline phases have been demonstrated theoretically and experimentally~\cite{open_DTC_1,open_DTC_2,open_DTC_3,open_DTC_4,open_DTC_5,open_DTC_6}. Relatedly, small interacting quantum systems have been shown to exhibit prethermal DTC behavior through the emergence of long-lived dynamical plateaus under spin-locking protocols~\cite{open_DTC_7}. In a similar spirit, an environment-assisted DTC has recently been demonstrated in a system of noninteracting two-level systems, where a separation between the transverse decoherence and longitudinal relaxation times allows environmental decoherence to suppress unwanted components of the dynamics while preserving the relevant magnetization~\cite{Das2026}. This mechanism, therefore,  may not require interactions, disorder, or MBL, and its theoretical description is naturally formulated in terms of the Liouvillian rather than a Floquet Hamiltonian. It therefore illustrates that time-crystalline behavior can emerge from dynamical mechanisms fundamentally different from eigenstate localization.

Open quantum systems can also be viewed from the perspective of quantum measurement~\cite{manzano2018thermodynamics}. Continuously monitored many-body systems have attracted considerable interest in recent years. A central finding is that they exhibit measurement-induced entanglement transitions, in which the entanglement of individual quantum trajectories changes from volume-law to area-law scaling as the measurement strength increases~\cite{Turkeshi2021,Ising_2,Ising_3,diff_unravelling_Ising,muller2025monitored,starchl2025generalized,Chatterjee2025,quasiperiodic_measurement}. In isolated driven systems, by contrast, extensive entanglement growth accompanies thermalization and destroys time-crystalline order; this is why nonergodicity, provided either by disorder-induced localization or by nonergodic Floquet eigenstates, has been central to stabilizing DTC phases~\cite{DTC_th_8,DTC_th_11,Bothra2025}. Measurements can also be used to condition the dynamics itself: selecting the most likely measurement record defines a deterministic yet non-unitary evolution, which has recently been used to study measurement-induced transitions~\cite{most_prob_trj_2026}, and partially postselected monitoring interpolates between this limit and ordinary quantum trajectories~\cite{partial_post_selection}. Stochastic measurements have also been shown to produce intermittent time-crystalline dynamics~\cite{DTC_projective}. A natural question therefore arises: can the low-entanglement regime generated by continuous measurements play a role analogous to localization, such that a DTC emerges on the area-law side of the measurement-induced entanglement transition?

In this work, we address this question in a clean, long-range interacting Floquet spin chain whose local occupations are continuously and weakly monitored. Our main findings are summarized in Fig.~\ref{fig:schematic}. We find that entanglement suppression alone is not sufficient: along typical stochastic trajectories, the subharmonic response decays even in the area-law regime, consistent with the featureless ensemble-averaged dynamics. In contrast, along the most-probable quantum trajectory~\cite{most_prob_trj_2026}, the deterministic and nonlinear measurement backaction stabilizes robust period doubling at sufficiently strong measurement, accompanied by suppressed entanglement and rigidity against pulse imperfections. To connect these two limits, we introduce a finite-width postselection protocol that bounds the measurement noise~\cite{partial_post_selection}. We show that period doubling survives for narrow noise windows, that intermediate windows produce two coexisting classes of trajectories, and that the order is progressively lost as the noise distribution approaches its unbounded Gaussian form with large-deviation trajectories.

The rest of the paper is organized as follows. In Sec.~\ref{sec:monitoring}, we review continuous weak measurements and introduce the finite-width postselection protocol together with its two limiting cases. In Sec.~\ref{sec:model}, we present the model and the drive protocol. Section~\ref{sec:observables} defines the initial states and the observables, including the subharmonic order parameter. Our numerical results are presented in Sec.~\ref{sec:results}, and Sec.~\ref{sec:conclusion} contains our conclusions and a discussion of experimental prospects. Additional data are collected in Appendixes~\ref{app:unbounded}--\ref{app:alpha}.

\section{Monitoring protocol}
\label{sec:monitoring}

We briefly review continuous weak measurements~\cite{measurement_def}. For a weak measurement of an observable $\hat x$ during a time $dt$, the measurement operator associated with an outcome $\beta$ is
\begin{equation*}
  \hat O(\beta)=\left(\frac{2\gamma\,dt}{\pi}\right)^{1/4} e^{-\gamma\,dt\,(\hat x-\beta)^2},
\end{equation*}
where $\gamma$ is the measurement strength. After the measurement, the state becomes $\hat O(\beta)\ket{\psi(t)}$, up to normalization. For $\gamma\,dt\ll1$, the outcome probability is
\begin{align*}
  p(\beta)&=\bra{\psi(t)}\hat O^\dagger(\beta)\hat O(\beta)\ket{\psi(t)}\\
  &\simeq\left(\frac{2\gamma\,dt}{\pi}\right)^{1/2} e^{-2\gamma\,dt\,(\beta-\langle\hat x\rangle_t)^2},
\end{align*}
where $\langle\hat x\rangle_t=\bra{\psi(t)}\hat x\ket{\psi(t)}$ is the expectation value in the current state. Thus, $\beta$ is Gaussian distributed with mean $\langle\hat x\rangle_t$ and variance $1/(4\gamma\,dt)$. One can write $\beta=\langle\hat x\rangle_t+dW/(2\gamma\,dt)$, where $dW$ is a Gaussian noise increment with $\avtraj{dW}=0$ and $\avtraj{dW^2}=\gamma\,dt$; here $\avtraj{\cdots}$ denotes the average over measurement trajectories [defined in Eq.~\eqref{eq:trajavg}]. The most probable outcome is $\beta=\langle\hat x\rangle_t$, that is, $dW=0$. Selecting this outcome at every step defines the most-probable trajectory~\cite{most_prob_trj_2026}.

We consider a many-body system with Hamiltonian $\hat H$, in which the local particle number $\hat n_l$ at each site $l$ is continuously monitored. Since the $\hat n_l$'s at different sites commute, the measurement operator factorizes as
\begin{equation*}
  \hat O(\{\beta_{l,t}\})=\prod_l\left(\frac{2\gamma\,dt}{\pi}\right)^{1/4} e^{-\gamma\,dt\,(\hat n_l-\beta_{l,t})^2},
\end{equation*}
with $\beta_{l,t}=\langle\hat n_l\rangle_t+dW_{l,t}/(2\gamma\,dt)$. The noise increments are independent, with $\avtraj{dW_{l,t}}=0$ and $\avtraj{dW_{l,t}\,dW_{l',t'}}=\gamma\,dt\,\delta_{ll'}\delta_{tt'}$. Including the unitary evolution, the normalized state after a time $dt$ is
\begin{equation*}
  \ket{\psi(t+dt)}\propto\hat O(\{\beta_{l,t}\})\,e^{-i\hat H dt}\ket{\psi(t)}.
\end{equation*}

To study fluctuations around the most-probable trajectory, we introduce a finite-width postselection protocol, similar to that of Ref.~\cite{partial_post_selection}. We accept only increments with $|dW_{l,t}|<w\sqrt{\gamma\,dt}$, where the dimensionless parameter $w$ sets the window width. The accepted increments follow the truncated Gaussian distribution
\begin{equation}
  P_w(dW_{l,t})=\frac{e^{-dW_{l,t}^2/(2\gamma\,dt)}}{\sqrt{2\pi\gamma\,dt}\;\mathrm{erf}(w/\sqrt2)},\qquad |dW_{l,t}|<w\sqrt{\gamma\,dt}.
  \label{eq:Pw}
\end{equation}
All trajectories are generated with this distribution, so that, for finite $w$, $\avtraj{\cdots}$ refers to the postselected ensemble. The noise moments become $\avtraj{dW_{l,t}}=0$ and $\avtraj{dW_{l,t}\,dW_{l',t'}}=\kappa(w)\,\gamma\,dt\,\delta_{ll'}\delta_{tt'}$, with
\begin{equation*}
  \kappa(w)=1-\frac{2w\,e^{-w^2/2}}{\sqrt{2\pi}\;\mathrm{erf}(w/\sqrt2)},
\end{equation*}
which increases monotonically from $\kappa\to0$ for $w\to0$ to $\kappa\to1$ for $w\to\infty$.

We expand the update to $O(dt)$, absorb constant terms into the normalization, and replace $dW_{l,t}\,dW_{l',t}$ by its average $\kappa(w)\,\gamma\,dt\,\delta_{ll'}$. With $\delta\hat n_l=\hat n_l-\langle\hat n_l\rangle_t$, this yields the finite-width stochastic Schr\"odinger equation (SSE)
\begin{align}
  d\ket{\psi}=\Big[&-i\hat H dt+\sum_l\delta\hat n_l\,dW_{l,t}
  +\gamma\,dt\Big(\frac{\kappa}{2}-1\Big)\sum_l\delta\hat n_l^2 \nonumber\\
  &-\gamma\,dt\,(\kappa-1)\sum_l\langle\delta\hat n_l^2\rangle_t\Big]\ket{\psi}.
  \label{eq:SSE}
\end{align}
In our simulations, we integrate Eq.~\eqref{eq:SSE} using the Trotterized form~\cite{cao2019entanglement}
\begin{equation}
  \ket{\psi(t+dt)}=\frac{1}{\mathcal N_P}\,e^{\hat M_P}\,e^{-i\hat H dt}\ket{\psi(t)},
  \label{eq:trotter}
\end{equation}
where $\mathcal N_P$ is the normalization and
\begin{equation*}
  \hat M_P=\sum_l\Big[dW_{l,t}+\big(2\langle\hat n_l\rangle_t-1\big)\gamma\,dt\Big]\hat n_l .
\end{equation*}
Here we have used $\hat n_l^2=\hat n_l$ for hard-core bosons. We study the protocol for general $w$ and also examine its two limiting cases separately:

(i) \emph{Most-probable deterministic protocol} ($w\to0$). The noise vanishes and $\kappa\to0$. The Trotterized update then contains no noise term, and the SSE becomes deterministic,
\begin{equation}
  d\ket{\psi}=\Big[-i\hat H dt-\gamma\,dt\sum_l\big(\delta\hat n_l^2-\langle\delta\hat n_l^2\rangle_t\big)\Big]\ket{\psi}.
  \label{eq:SSE_mp}
\end{equation}
This evolution is deterministic but remains nonlinear and state dependent through $\langle\hat n_l\rangle_t$.

(ii) \emph{Unbounded stochastic protocol} ($w\to\infty$). The truncation is removed allowing large-deviation trajectories and $\kappa\to1$. All trajectories are sampled with Born-rule probabilities, and the SSE reduces to the standard form
\begin{equation}
  d\ket{\psi}=\Big[-i\hat H dt+\sum_l\delta\hat n_l\,dW_{l,t}-\frac{\gamma\,dt}{2}\sum_l\delta\hat n_l^2\Big]\ket{\psi}.
  \label{eq:SSE_mp1}
\end{equation}

The width $w$ thus interpolates between the most-probable trajectory and ordinary quantum trajectories. By varying $w$, we study how fluctuations around the most-probable measurement record affect the period-doubled response.

\section{Model and drive protocol}
\label{sec:model}

We consider a clean, long-range interacting spin-$1/2$ chain of $L$ sites and ask whether continuous weak measurements can stabilize a DTC without disorder. The monitored observable is the local occupation $\hat n_l=(1+\sigma^z_l)/2$, which follows from mapping each spin to a hard-core boson, with spin up (down) corresponding to an occupied (empty) site. Monitoring $\hat n_l$ is thus equivalent to monitoring $\sigma^z_l$.

We use a two-step Floquet protocol. The first step is a global spin rotation about the $x$ axis. The second step combines a long-range Ising interaction with a nearest-neighbor $XY$ coupling, and the measurements act only during this second step. A similar model without measurements was studied in Ref.~\cite{Bothra2025}, where the DTC was stabilized by disorder. The Hamiltonians of the two steps are
\begin{align}
  H_1&=\Big(\frac{\pi}{2}-\epsilon\Big)\sum_l\sigma^x_l, \nonumber\\
  H_2&=J\sum_{l<m}\frac{\sigma^z_l\sigma^z_m}{|l-m|^\alpha}
  +g\sum_l\big(\sigma^x_l\sigma^x_{l+1}+\sigma^y_l\sigma^y_{l+1}\big).
  \label{eq:H}
\end{align}
We use open boundary conditions, so that the $XY$ sum runs over $l=1,\dots,L-1$. One Floquet period has duration $T=2$. Starting from $\ket{\psi_{\mathrm{ini}}}$, the first step gives $\ket{\psi(T/2)}=e^{-iH_1T/2}\ket{\psi_{\mathrm{ini}}}$. Since $T/2=1$, this rotates every spin about the $x$ axis by an angle $\pi-2\epsilon$. For $\epsilon=0$, it is a perfect spin flip, $e^{-iH_1}=\prod_l(-i\sigma^x_l)$, while a nonzero $\epsilon$ makes the flip imperfect. We consider $\alpha=1$ unless otherwise specified.

During the second step, the system evolves under $H_2$ and is monitored as described in Sec.~\ref{sec:monitoring}. We divide this step into $N_t$ intervals of length $dt$, with $N_t\,dt=T/2$, and write $t_j=T/2+j\,dt$. Each interval applies Eq.~\eqref{eq:trotter} with $\hat H=H_2$,
\begin{equation}
  \ket{\psi(t_{j+1})}=\frac{1}{\mathcal N_P}\,e^{\hat M_P(t_j)}\,e^{-iH_2dt}\ket{\psi(t_j)},
  \label{eq:step}
\end{equation}
for $j=0,\dots,N_t-1$. Here, $\langle\hat n_l\rangle_{t_j}$ in $\hat M_P(t_j)$ is evaluated in $\ket{\psi(t_j)}$, and the increments $dW_{l,t_j}$ are drawn from $P_w$ in Eq.~\eqref{eq:Pw}. Iterating Eq.~\eqref{eq:step} gives $\ket{\psi(T)}$, and the Floquet sequence is then repeated.

This scheme converges to the continuous-time dynamics as $dt\to0$. We use $dt=0.05$ throughout and have verified that the results do not change appreciably for smaller $dt$. 
We use $J=2$, $g=0.5$, and $\alpha=1$ throughout the main text. In Appendix~\ref{app:alpha}, we show results for $\alpha=2$, which confirm that our findings are robust to the range of the interactions.

\begin{figure*}[t]
  \centering
  \includegraphics[width=\textwidth]{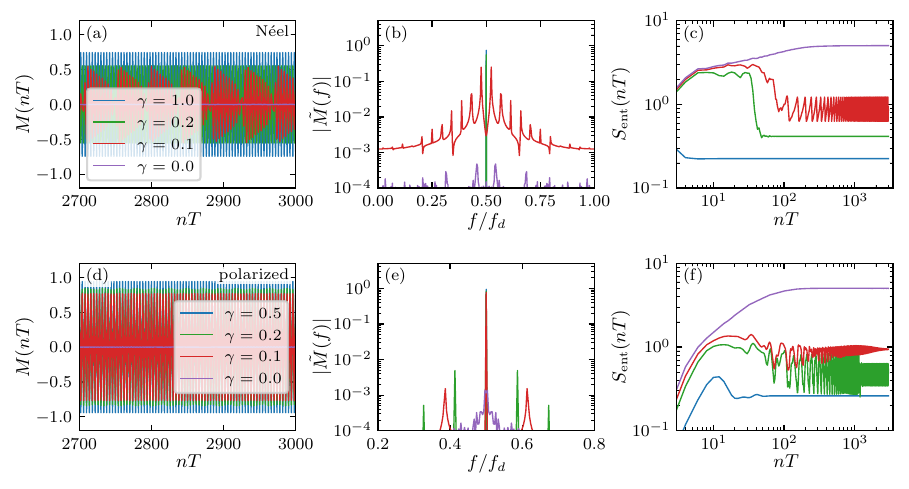}
  \caption{Most-probable-trajectory ($w=0$) dynamics for (a)--(c) the N\'eel state and (d)--(f) the fully polarized state. (a), (d) Stroboscopic magnetization $M(nT)$ [Eq.~\eqref{eq:M}; staggered for the N\'eel state] at late times. (b), (e) Fourier power spectrum $|\tilde M(f)|$ of the late-time signal [Eq.~\eqref{eq:fourier}]. (c), (f) Half-chain entanglement entropy $\Sent(nT)$. Colors denote measurement strengths $\gamma$ increasing from purple (weakest) via red, green, and orange to blue (strongest). Fourier spectra are computed in a window $n \in [1200,1500]$. Results are for $L=16$.}
  \label{fig:mp_dynamics}
\end{figure*}

\section{Initial states and observables}
\label{sec:observables}

We consider two initial product states in the occupation basis, where 1 (0) denotes an occupied (empty) site: the fully polarized state $\ket{\psi_P}=\ket{111\cdots1}$ and the N\'eel state $\ket{\psi_N}=\ket{0101\cdots}$. Both are simple to prepare in experiments.

\emph{Averages.} Each realization of the measurement record defines a quantum trajectory $\ket{\psi^{(r)}(t)}$, labeled by $r=1,\dots,N_{\mathrm{traj}}$. All observables below are first evaluated within each trajectory, using the local expectation values $\langle\sigma^z_l\rangle^{(r)}_t=\bra{\psi^{(r)}(t)}\sigma^z_l\ket{\psi^{(r)}(t)}$ or the reduced density matrix of $\ket{\psi^{(r)}(t)}$. Denoting such a single-trajectory quantity by $F^{(r)}$, its trajectory average is
\begin{equation}
  \avtraj{F}=\frac{1}{N_{\mathrm{traj}}}\sum_{r=1}^{N_{\mathrm{traj}}}F^{(r)},
  \label{eq:trajavg}
\end{equation}
taken over the ensemble generated with the noise distribution $P_w$ of Eq.~\eqref{eq:Pw}. In the limit ($w\to0$), almost all trajectories coincide, and the trajectory-averaged quantity coincides with individual trajectory values. In the most-probable trajectory limit, $w=0$, we have $\avtraj{F}=F^{(1)}$. In addition, an overline denotes the late-time stroboscopic average,
\begin{equation}
  \overline F=\frac{1}{N_{\mathrm{late}}}\sum_{n=n_0}^{n_{\mathrm{end}}}F(nT),
  \label{eq:lateavg}
\end{equation}
where the averaging window runs over the Floquet periods $n_0\le n\le n_{\mathrm{end}}$, with $n_0$ chosen after transients have decayed (its value depends on the parameters and the initial state), and $N_{\mathrm{late}}=n_{\mathrm{end}}-n_0+1$ is the number of periods in the window. Throughout, we label late-time windows by their last period $n_{\mathrm{end}}$.

\emph{Magnetization and subharmonic response.} To detect the DTC, we measure the magnetization relative to the initial spin pattern. For $r$-th trajectory,
\begin{equation}
  M^{(r)}(t)=\frac1L\sum_{l=1}^{L}s_l\,\langle\sigma^z_l\rangle^{(r)}_t,
  \label{eq:M}
\end{equation}
with $s_l=(-1)^{2l}=1~ \forall ~ l$  [$s_l=(-1)^{l}=-1,1, -1,1,...$ for $l=1,2,3,4,...$] for initial polarized [N\'eel] state $\ket{\psi_P}$ [$\ket{\psi_N}$].  
We use the single symbol $M(t)=\avtraj{M(t)}$, after trajectory averaging for both the initial states, and $M(0)=1$ in both cases. The period-doubled response of trajectory $r$ is quantified by the subharmonic response, defined with the late-time average of Eq.~\eqref{eq:lateavg},
\begin{equation}
  m^{(r)}_{\mathrm{sub}}=\Big|\,\overline{(-1)^nM(^{(r)}(nT)}\,\Big|.
  \label{eq:msub}
\end{equation}
In a DTC, the stroboscopic magnetization oscillates with period $2T$, $M^{(r)}(nT)\simeq A(-1)^n$, where $A$ is a constant amplitude with $0<A\le1$, and Eq.~\eqref{eq:msub} gives $m^{(r)}_{\mathrm{sub}}\simeq A$. For perfect period doubling, $A=1$, while an imperfect DTC has $A<1$. Without a DTC, the period-$2T$ oscillation decays, and $m^{(r)}_{\mathrm{sub}}\to0$. Equivalently, the period-doubled response is characterized by the recurrence of the magnetization at even multiples of the driving period, $M(2nT)\simeq A$, together with its sign reversal at odd multiples, while the finite subharmonic response $m_{\mathrm{sub}}\simeq A>0$ quantifies the persistent period doubling. We characterize the ensemble by the distribution $p(\msub)$ of the values $m^{(r)}_{\mathrm{sub}}$ and by their mean $\avtraj{\msub}$. For $w=0$, there is a single trajectory and we simply write $\msub$.

\emph{Fourier spectrum.} To further characterize the response, we compute the Fourier spectrum of the stroboscopic magnetization in the late-time window,
\begin{equation}
  \tilde M^{(r)}(f)=\left|\frac{1}{N_{\mathrm{late}}}\sum_{n=n_0}^{n_{\mathrm{end}}}M^{(r)}(nT)\,e^{-2\pi i f nT}\right|,
  \label{eq:fourier}
\end{equation}
and its trajectory average $\avtraj{\tilde M(f)}$, where $f_d=1/T$ is the drive frequency. Period doubling appears as a peak at $f/f_d=1/2$.

\emph{Entanglement entropy.} Continuously monitored systems can also show an entanglement transition from volume-law to area-law scaling. To compare this transition with the DTC order, we divide the chain into a subsystem $A$, consisting of the first $l_A=L/2$ sites, and its complement $B$. For trajectory $r$, the reduced density matrix of $A$ is $\rho^{(r)}_A(t)=\mathrm{Tr}_B\ket{\psi^{(r)}(t)}\bra{\psi^{(r)}(t)}$, and the half-chain entanglement entropy is
\begin{equation*}
  \Sent^{(r)}(t)=-\mathrm{Tr}_A\big[\rho^{(r)}_A(t)\ln\rho^{(r)}_A(t)\big].
\end{equation*}
We average it over trajectories, $\avtraj{\Sent(t)}$, and over the late-time window, $\overline{\avtraj{\Sent}}$. In the area-law phase, $\overline{\avtraj{\Sent}}$ remains fixed as $L$ increases, whereas in the volume-law phase it grows linearly with $L$.

Note that $M(t)$ is linear in the state, so it can always be written as an expectation value in the trajectory-averaged density matrix $\rho(t)=\avtraj{\ket{\psi(t)}\bra{\psi(t)}}$. In the unbounded protocol ($w\to\infty$), $\rho(t)$ moreover obeys a closed Lindblad master equation, so $M(t)$ is fully determined by this Lindbladian dynamics (Sec.~\ref{sec:res_unbounded}). For finite $w$, the postselection makes the averaged dynamics nonlinear, and $\rho(t)$ no longer follows a Lindblad equation. In all cases, $m^{(r)}_{\mathrm{sub}}$ and $\Sent^{(r)}$ are evaluated in individual trajectories before averaging and therefore probe properties of the trajectories that are not captured by $\rho(t)$.

\section{Results}
\label{sec:results}

We now present numerical results for three measurement protocols: (i) the unbounded stochastic protocol ($w\to\infty$), (ii) the most-probable deterministic protocol ($w=0$), and (iii) the bounded stochastic protocol ($0<w<\infty$).

\subsection{Unbounded stochastic protocol ($w\to\infty$)}
\label{sec:res_unbounded}

In this limit, the postselection window is removed, so all measurement outcomes are accepted. The noise increments $dW_{l,t}$ then follow the full Gaussian distribution, and outcomes are sampled with Born-rule probabilities. Each measurement record generates a different quantum trajectory; we compute all quantities for each trajectory and then average over trajectories.

As noted in Sec.~\ref{sec:observables}, the magnetization is linear in the state, so its trajectory average is fixed by the ensemble-averaged density matrix $\rho(t)=\avtraj{\ket{\psi(t)}\bra{\psi(t)}}$, which obeys a Lindblad master equation with dephasing operators $\hat n_l$. This dynamics heats the system to the featureless infinite-temperature state, $\rho\propto\hat{\mathbb I}$~\cite{cao2019entanglement,steady_state_lindblad}. In this state, $\mathrm{Tr}[\rho\,\sigma^z_l]=0$, so the trajectory-averaged magnetization $M$ and its subharmonic component $|\overline{(-1)^nM}|$ vanish at late times. We therefore do not expect a DTC in this limit, and our numerics confirm this for all values of $\gamma$ that we have studied (Appendix~\ref{app:unbounded}).

The entanglement entropy, in contrast, is a nonlinear function of the state. It is computed for each pure-state trajectory before averaging and is therefore not fixed by $\rho(t)$: individual trajectories can remain weakly or strongly entangled even though their average state is featureless. As $\gamma$ increases, $\overline{\avtraj{\Sent}}$ indeed shows a transition from volume-law to area-law scaling, consistent with measurement-induced transitions in related systems~\cite{Turkeshi2021,Ising_2,Ising_3,diff_unravelling_Ising,muller2025monitored,starchl2025generalized} (Fig.~\ref{fig:unbounded} in Appendix~\ref{app:unbounded}). Thus, there is no one-to-one correspondence between the entanglement transition and DTC order; in particular, area-law entanglement does not imply a DTC.

\begin{figure}[t]
  \centering
  \includegraphics[width=\columnwidth]{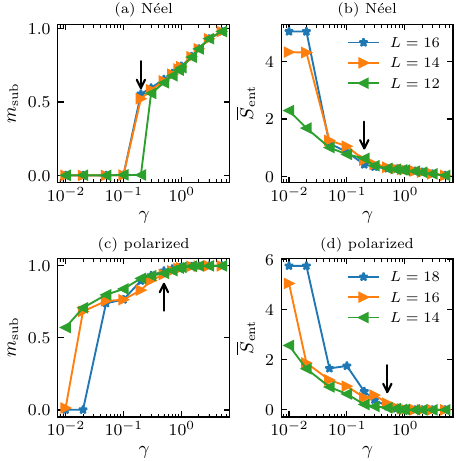}
  \caption{Most-probable trajectory ($w=0$): (a), (c) subharmonic response $\msub$ and (b), (d) late-time entanglement entropy $\overline{\Sent}$ versus $\gamma$ for (a), (b) the N\'eel state with $L=12$, 14, 16 and (c), (d) the fully polarized state with $L=14$, 16, 18. Arrows mark the onset of period doubling. Parameters: $\epsilon=0.15$, $J=2$, $g=0.5$, $\alpha=1$. Late-time averaging is done in the window $n \in [1200,1500]$.}
  \label{fig:mp_gamma}
\end{figure}

\begin{figure}[t]
  \centering
  \includegraphics[width=1\columnwidth,height=0.98\columnwidth]{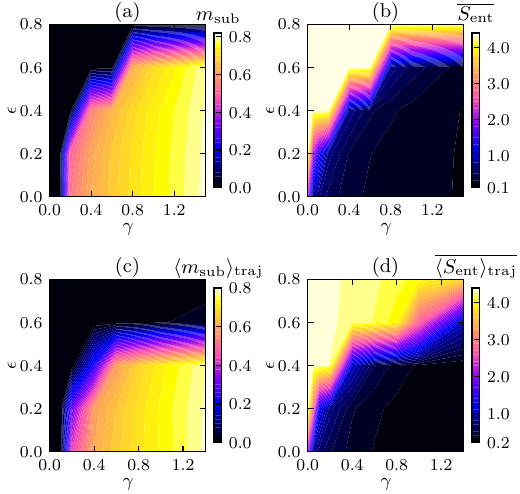}
 \caption{Phase diagrams for the N\'eel state in the $(\gamma,\epsilon)$ plane. (a) Subharmonic response $\msub$ and (b) late-time entanglement entropy $\overline{\Sent}$ along the most-probable trajectory ($w=0$). (c) Mean subharmonic response $\avtraj{\msub}$ and (d) late-time entanglement entropy $\overline{\avtraj{\Sent}}$ for the bounded protocol with $w=0.3$. Parameters: $L=14$, $J=2$, $g=0.5$, $\alpha=1$. Late-time averaging is done in the window $n \in [1200,1500]$.}
  \label{fig:phase}
\end{figure}

\subsection{Most-probable deterministic protocol ($w=0$)}
\label{sec:res_mp}

We now set the noise to zero, $dW_{l,t}=0$. The evolution, Eq.~\eqref{eq:SSE_mp}, is then deterministic, but the measurement backaction keeps it nonlinear and state dependent. We first discuss the dynamics, then the dependence on $\gamma$ and $L$, and finally the robustness against $\epsilon$. Since this protocol generates a single trajectory, no trajectory average is needed [Eq.~\eqref{eq:trajavg}], and we write simply $M$, $\msub$, $\Sent$, and $\overline{\Sent}$.

\subsubsection{Dynamics}

Figure~\ref{fig:mp_dynamics} shows the dynamics for $L=16$ and $\epsilon=0.15$. For the N\'eel state [Figs.~\ref{fig:mp_dynamics}(a)--\ref{fig:mp_dynamics}(c)], weak measurement does not produce a persistent period-doubled oscillation: the spectrum has no dominant peak at $f/f_d=1/2$, and the entanglement entropy grows to a large value. At intermediate $\gamma$, the spectrum shows several peaks around $f/f_d=1/2$. These indicate beating between competing frequencies, which is also visible as oscillations in $\Sent(nT)$. At strong measurement, a single sharp peak at $f/f_d=1/2$ dominates the spectrum, and at the same time the entanglement entropy saturates at a small value.

The fully polarized state [Figs.~\ref{fig:mp_dynamics}(d)--\ref{fig:mp_dynamics}(f)] shows the same crossover, but at much weaker measurement. Already at intermediate $\gamma$, $M(nT)$ oscillates with period $2T$ and the spectrum has a single peak at $f/f_d=1/2$. As $\gamma$ increases, the oscillation amplitude approaches unity, and the entanglement is strongly suppressed.

\begin{figure*}[t]
  \centering
  \includegraphics[width=0.92\textwidth]{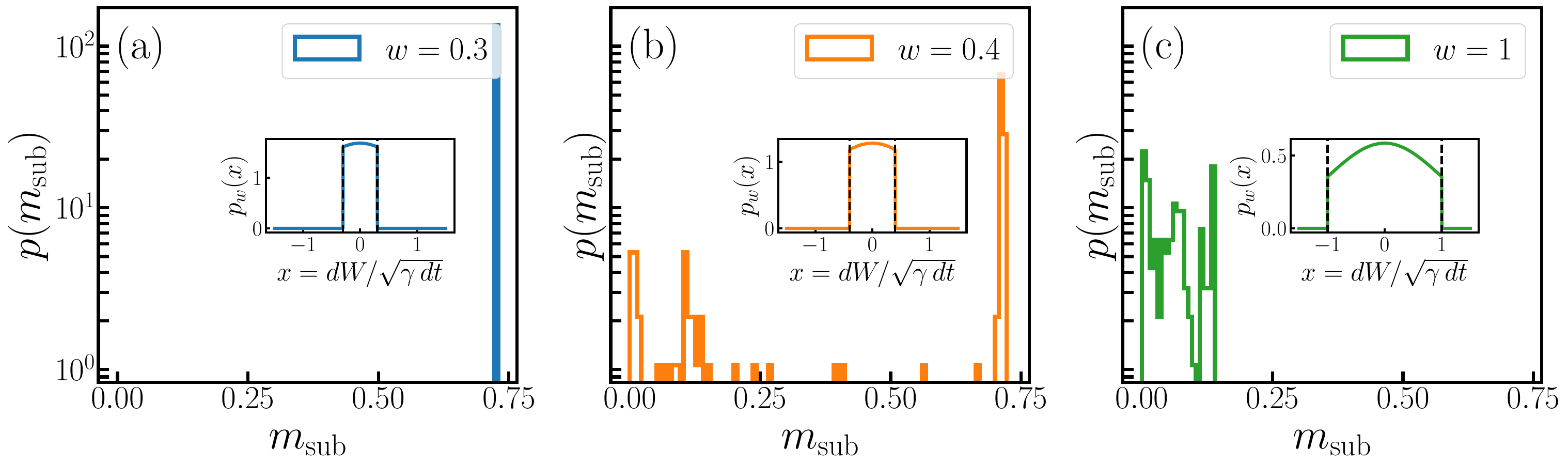}
  \caption{Histogram $p(\msub)$ of the late-time subharmonic response over trajectories for (a) $w=0.3$, (b) $w=0.4$, and (c) $w=1$. Insets: truncated noise distribution $p_w(x)$ [Eq.~\eqref{eq:Pw}] of $x=dW/\sqrt{\gamma\,dt}$; dashed lines mark the window $|x|<w$. Measurement action is incorporated, followed by Eq.~\eqref{eq:trotter}. Data is for N\'eel state, $L=14$, $\epsilon=0.15$, $\gamma=1$, $n_0=1200, ~n_{\mathrm{end}}=1500$.}
  \label{fig:hist}
\end{figure*}

\begin{figure*}[t]
  \centering
  \includegraphics[width=0.92\textwidth]{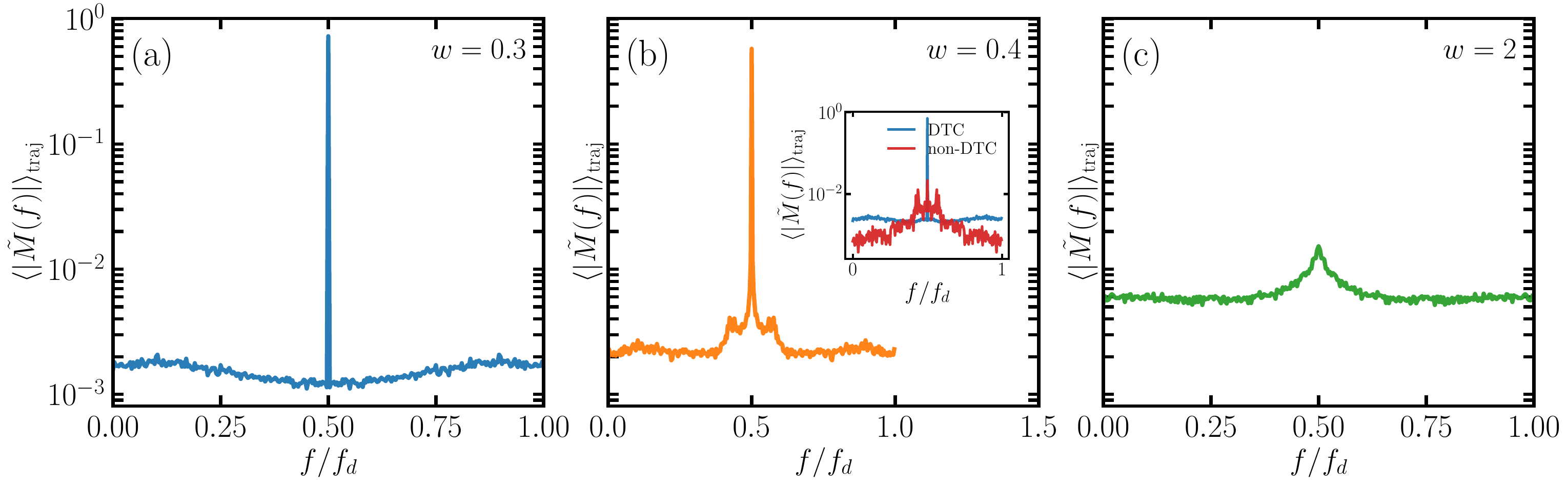}
  \caption{Trajectory-averaged Fourier spectra $\avtraj{\tilde M(f)}$ [Eq.~\eqref{eq:fourier}] of the late-time stroboscopic magnetization for (a) $w=0.3$, (b) $w=0.4$, and (c) $w=2$. Inset of (b): spectra at $w=0.4$ averaged separately over the time-crystalline (DTC) trajectories with $m^{(r)}_{\mathrm{sub}}>0.65$ (84 trajectories, blue) and the non-time-crystalline (non-DTC) trajectories with $m^{(r)}_{\mathrm{sub}}<0.1$ (14 trajectories, red). Data is for N\'eel state, $L=14$, $\epsilon=0.15$, $\gamma=1$, $n_0=1200,~n_{\mathrm{end}}=1500$.}
  \label{fig:spectra}
\end{figure*}

\subsubsection{Dependence on measurement strength and system size}

Figure~\ref{fig:mp_gamma} shows $\msub$ and $\overline{\Sent}$ as functions of $\gamma$ for several $L$. For the N\'eel state, $\msub$ vanishes for $\gamma\lesssim0.2$. It then jumps to about $0.5$ and grows continuously, approaching unity at large $\gamma$. The onset of period doubling coincides with a sharp drop of the entanglement entropy. Below the onset, the entropy grows with $L$, whereas above it the entropy no longer depends on $L$, in agreement with area-law scaling. For the polarized state, the onset occurs about an order of magnitude earlier, at $\gamma\approx0.01$--$0.05$ depending on $L$, together with a rapid suppression of the entanglement; $\msub$ then saturates close to unity for $\gamma\gtrsim1$.

For both states, the curves for different $L$ nearly collapse above the onset. The period doubling therefore persists for all accessible system sizes. We have also checked that the running average of $\msub$ saturates as a function of the number of periods, showing that the response does not decay over the simulated time.

\subsubsection{Robustness against pulse imperfections}

A DTC must be robust against small changes in the drive. We therefore vary the pulse imperfection $\epsilon$ for the N\'eel state. Figures~\ref{fig:phase}(a) and \ref{fig:phase}(b) show $\msub$ and $\overline{\Sent}$ in the $(\gamma,\epsilon)$ plane. At weak measurement, $\msub$ vanishes for all $\epsilon$, and the entanglement entropy is large. Increasing $\gamma$ produces a broad region with large $\msub$, in which the entanglement is strongly suppressed. The boundary of this region moves to larger $\gamma$ as $\epsilon$ increases: a larger pulse error requires a stronger measurement to stabilize period doubling. Within the DTC region, $\msub$ is nearly independent of $\epsilon$. The response is thus rigid against pulse imperfections, which is a hallmark of a DTC. Cuts at fixed $\epsilon$ are shown in Appendix~\ref{app:epsilon}.

Comparing the two maps, period doubling appears only where the entanglement is low: the DTC region in Fig.~\ref{fig:phase}(a) closely follows the low-entropy region in Fig.~\ref{fig:phase}(b), and the boundary of both moves to larger $\gamma$ with increasing $\epsilon$. Low entanglement is thus necessary for the DTC along the most-probable trajectory. It is, however, not sufficient in general: in the unbounded protocol, area-law entanglement appears without any period doubling (Sec.~\ref{sec:res_unbounded}).

\begin{figure}[t]
  \centering
  \includegraphics[width=\columnwidth]{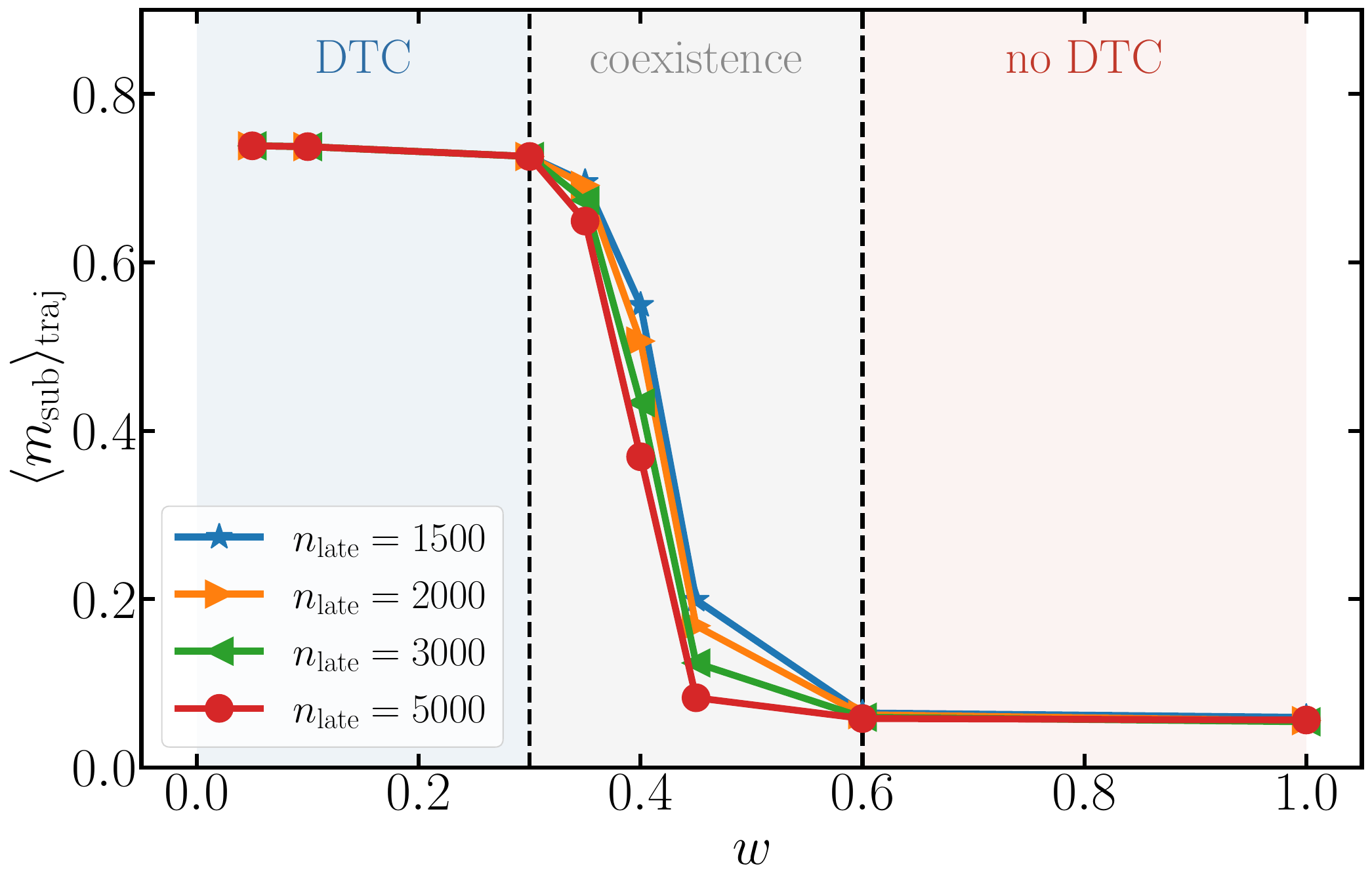}
  \caption{Mean subharmonic response $\avtraj{\msub}$ as a function of the postselection width $w$ for late-time windows ending at $n_{\mathrm{end}}=1500$, 2000, 3000, and 5000, with $n_0=1200$. Dashed lines at $w=0.3$ and $w=0.6$ separate the three regimes, shaded as in Fig.~\ref{fig:schematic}(d): all trajectories time crystalline (DTC, $w\le0.3$), coexistence of DTC and non-DTC trajectories ($0.3<w<0.6$), and no DTC ($w\ge0.6$). Parameters as in Fig.~\ref{fig:hist}.}
  \label{fig:nend}
\end{figure}

\subsection{Bounded stochastic protocol ($0<w<\infty$)}
\label{sec:res_bounded}

We now allow measurement noise within the window $|dW_{l,t}|<w\sqrt{\gamma\,dt}$. This protocol interpolates between the most-probable trajectory ($w\to0$) and the unbounded protocol with large-deviation trajectory ($w\to\infty$). Since trajectories now differ, we evaluate the subharmonic response $m^{(r)}_{\mathrm{sub}}$ of each trajectory [Eq.~\eqref{eq:msub}] and study its distribution $p(\msub)$ and mean $\avtraj{\msub}$. All results in this section are for the N\'eel state with $L=14$, averaged over $128$ trajectories.

\paragraph*{Phase diagram at finite $w$.}
Figures~\ref{fig:phase}(c) and \ref{fig:phase}(d) show $\avtraj{\msub}$ and $\overline{\avtraj{\Sent}}$ in the $(\gamma,\epsilon)$ plane for $w=0.3$. A broad DTC region survives. Compared with the most-probable trajectory [Fig.~\ref{fig:phase}(a)], two changes appear. The onset moves to a slightly larger measurement strength, $\gamma\approx0.3$, and the DTC region shrinks to $\epsilon\lesssim0.5$. Finite noise thus requires a stronger measurement and tolerates smaller pulse errors. As for $w\to0$, the DTC region lies inside the region of low entanglement. The period-doubled response therefore persists for finite noise, and it remains tied to suppressed entanglement.

\paragraph*{Histograms.}
To see how individual trajectories behave, we fix $\epsilon=0.15$ and $\gamma=1$, deep inside the DTC region. Figure~\ref{fig:hist} shows $p(\msub)$ for three values of $w$, with insets showing the corresponding truncated noise distributions. For $w=0.3$, all trajectories have $\msub\approx0.72$, almost the value of the most-probable trajectory. For $w=1$, all trajectories have $\msub\lesssim0.15$. For the intermediate value $w=0.4$, the histogram is nearly bimodal: most trajectories have $\msub\approx0.72$, while a smaller group has $\msub\lesssim0.15$ and a few trajectories take intermediate values. For a shorter-ranged Ising coupling ($\alpha=2$), the full evolution of $p(\msub)$ with $w$ is shown in Appendix~\ref{app:alpha}; there, the critical window $w_c$ below which the DTC survives is larger, $w_c\approx0.6$, compared with $w_c\approx0.3$--$0.4$ for $\alpha=1$.

\paragraph*{Spectra and two classes of trajectories.}
The trajectory-averaged spectra in Fig.~\ref{fig:spectra} show the same trend. Here we compute $\tilde M^{(r)}(f)$ [Eq.~\eqref{eq:fourier}] in each trajectory and then average over trajectories. For $w=0.3$ [Fig.~\ref{fig:spectra}(a)], the spectrum has a single sharp peak at $f/f_d=1/2$ on a low, smooth background. For $w=2$ [Fig.~\ref{fig:spectra}(c)], only a weak maximum remains on a broad background. For $w=0.4$ [Fig.~\ref{fig:spectra}(b)], the peak at $f/f_d=1/2$ is still strong, but it is accompanied by secondary peaks at $f/f_d\approx0.43$ and $0.57$.

To identify the origin of these secondary peaks, we split the trajectories according to their value of $m^{(r)}_{\mathrm{sub}}$ and average the spectra within each group [inset of Fig.~\ref{fig:spectra}(b)]. The 84 trajectories with $\msub>0.65$ show a single sharp primary peak at $f/f_d=1/2$ without side peaks, as for the most-probable trajectory; these trajectories support a time-crystalline phase. The 14 trajectories with $\msub<0.1$ show a much weaker peak at $f/f_d=1/2$, together with the side peaks at $f/f_d\approx0.43$ and $0.57$ and an elevated background; these trajectories have lost the period-doubled response. The secondary peaks in the averaged spectrum thus come entirely from this second group. At intermediate $w$, the averaged response is therefore a mixture of two distinct classes of trajectories, a structure that the mean value alone would hide. On the other hand, Fig.~\ref{fig:spectra}(c) shows a substantially diminished primary peak at $f/f_d=1/2$ signaling the disappearance of DTC phase for $\omega=2$.  

\paragraph*{Long-lived but finite DTC.}
To probe the stability over longer times, we extend the simulations up to 5000 Floquet periods. Figure~\ref{fig:nend} shows $\avtraj{\msub}$ as a function of $w$ for late-time windows ending at different $n_{\mathrm{end}}$. For $w\le0.3$, the curves coincide and stay at the most-probable value. For $w\gtrsim0.6$, they are close to zero. In between, $\avtraj{\msub}$ decreases with increasing $n_{\mathrm{end}}$, and the drop becomes sharper. In this intermediate regime, the time-crystalline trajectories are long lived, but they lose their order at later times.

The histograms in Appendix~\ref{app:hist_time} confirm this picture. At $w=0.4$, the peak at $\msub\approx0.72$ decreases steadily with increasing $n_{\mathrm{end}}$, while the weight at small $\msub$ grows; the position of the peak does not change. Individual trajectories therefore do not lose their amplitude gradually. Instead, they switch from the time-crystalline class to the non-time-crystalline class.

In summary, exact elimination of the measurement noise is not required. For narrow windows, all trajectories remain time crystalline over the simulated times. For intermediate windows, the DTC is long lived but eventually decays, as trajectories leave the time-crystalline class one by one.

\section{Conclusion}
\label{sec:conclusion}

In this work, we have investigated the emergence of subharmonic dynamics in a clean, long-range interacting Floquet spin chain subject to continuous quantum measurements. In contrast to conventional routes to discrete time-crystalline order, our system contains no disorder. We have focused on the dynamics of individual quantum trajectories and, in particular, on the most-probable trajectory, obtained by setting the stochastic measurement fluctuations to zero. This yields an evolution that is deterministic, yet non-unitary, nonlinear and state dependent owing to the measurement backaction.

Our central result concerns the relationship between entanglement and time-crystalline order in monitored systems. Across all protocols, we find that low entanglement alone is not sufficient to stabilize a DTC: in the unbounded protocol, typical trajectories display area-law entanglement at strong measurement, yet the subharmonic response vanishes. Conversely, whenever a stable DTC is present, the entanglement is low and obeys area-law scaling. Area-law entanglement is therefore a necessary, but not sufficient, condition for measurement-stabilized discrete time-crystalline order.

To reach this conclusion, we first considered the trajectory-averaged dynamics, obtained by averaging over different realizations of the measurement record. In the present system, the ensemble-averaged state approaches a featureless, infinite-temperature state at long times. Consequently, the magnetization decays and no persistent subharmonic response survives at the trajectory-averaged level. Notably, this remains true even in the regime where continuous measurements strongly suppress the entanglement and drive the system into an area-law phase.

Next, we studied the most-probable trajectory, in which measurement fluctuations are absent but the nonlinear backaction remains. For both N\'eel and fully polarized initial states, increasing the measurement strength drives the system into a low-entanglement regime with a robust period-doubled response, signaled by a sharp peak at $f/f_d=1/2$. The onset of this response depends on the initial state, occurring at weaker measurement for the fully polarized state, and persists across all accessible system sizes. The period doubling is also robust against small pulse imperfections $\epsilon$, with larger $\epsilon$ requiring stronger measurement.

Finally, to bridge the two limiting cases, we considered a bounded stochastic protocol ($0<w<\infty$), in which the measurement noise is retained but restricted to a finite window of width $w$. This protocol interpolates continuously between the deterministic most-probable trajectory ($w\to0$) and the unbounded stochastic dynamics ($w\to\infty$). We find that the period-doubled response survives for a finite range of $w$, demonstrating that exact elimination of the measurement noise is not required: it is sufficient to suppress large fluctuations of the measurement record. This range of tolerable noise widens when the interactions become shorter ranged (larger $\alpha$; Appendix~\ref{app:alpha}). At intermediate $w$, the ensemble separates into time-crystalline and non-time-crystalline trajectories, and the time-crystalline fraction decreases slowly in time as trajectories switch class one by one.

The bounded stochastic protocol is also particularly relevant for experiments. Accessing the exact most-probable trajectory would require postselecting a single measurement record, an event of vanishing probability, and even reconstructing individual trajectories generally incurs a postselection overhead that grows exponentially with system size and evolution time. In contrast, the bounded protocol only requires discarding runs in which the measurement signal leaves a finite acceptance window. A finite fraction of outcomes is then accepted at each time step, which reduces the overhead compared with selecting a single record, although the overall acceptance probability of a full run still decreases with system size and evolution time. Such conditioning on the measurement record is naturally compatible with continuous weak-measurement schemes available in trapped-ion and superconducting-qubit platforms, where long-range interactions and high-fidelity continuous readout have already been demonstrated. We therefore expect the bounded-noise protocol to provide a practical route to observing measurement-stabilized time-crystalline order in near-term quantum simulators. 
\begin{acknowledgments}
R.M. acknowledges the DST-INSPIRE fellowship from the Department of Science and Technology, Government of India. T.N. acknowledges the Advanced Research Grant (ARG) from
Anusandhan National Research Foundation Grant No.
ANRF/ARG/2025/003163/PS. Generative AI tools (Anthropic’s Claude) were used for editorial polishing and assistance. All scientific content,
analyses, interpretations, and conclusions were developed
and verified by the authors, who take full responsibility
for the manuscript.
\end{acknowledgments}

\section*{Data availability}
The data that support the findings of this article are available from the authors upon reasonable request.

\appendix

\begin{figure}[t]
  \centering
  \includegraphics[width=\columnwidth]{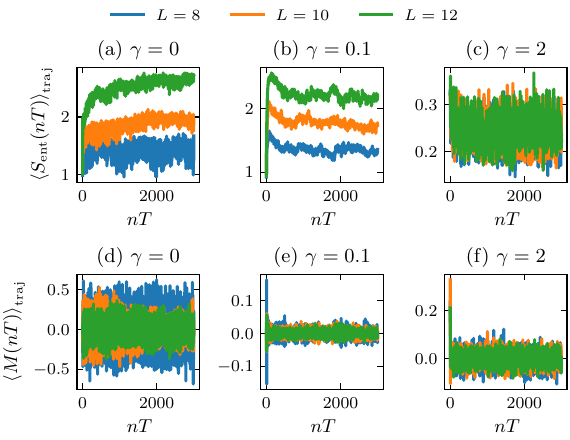}
  \caption{Unbounded stochastic protocol ($w\to\infty$). (a)--(c) Trajectory-averaged entanglement entropy $\avtraj{\Sent(nT)}$ and (d)--(f) trajectory-averaged stroboscopic magnetization $M(nT)$ for $\gamma=0$, 0.1, and 2 and system sizes $L=8$, 10, and 12. The initial state is the N\'eel state $\ket{\psi_N}$. Other parameters: $\epsilon=0.05$, $J=2$, $g=0.5$, $\alpha=1$; averages over 128 trajectories.}
  \label{fig:unbounded}
\end{figure}

\begin{figure}[t]
  \centering
  \includegraphics[width=\columnwidth]{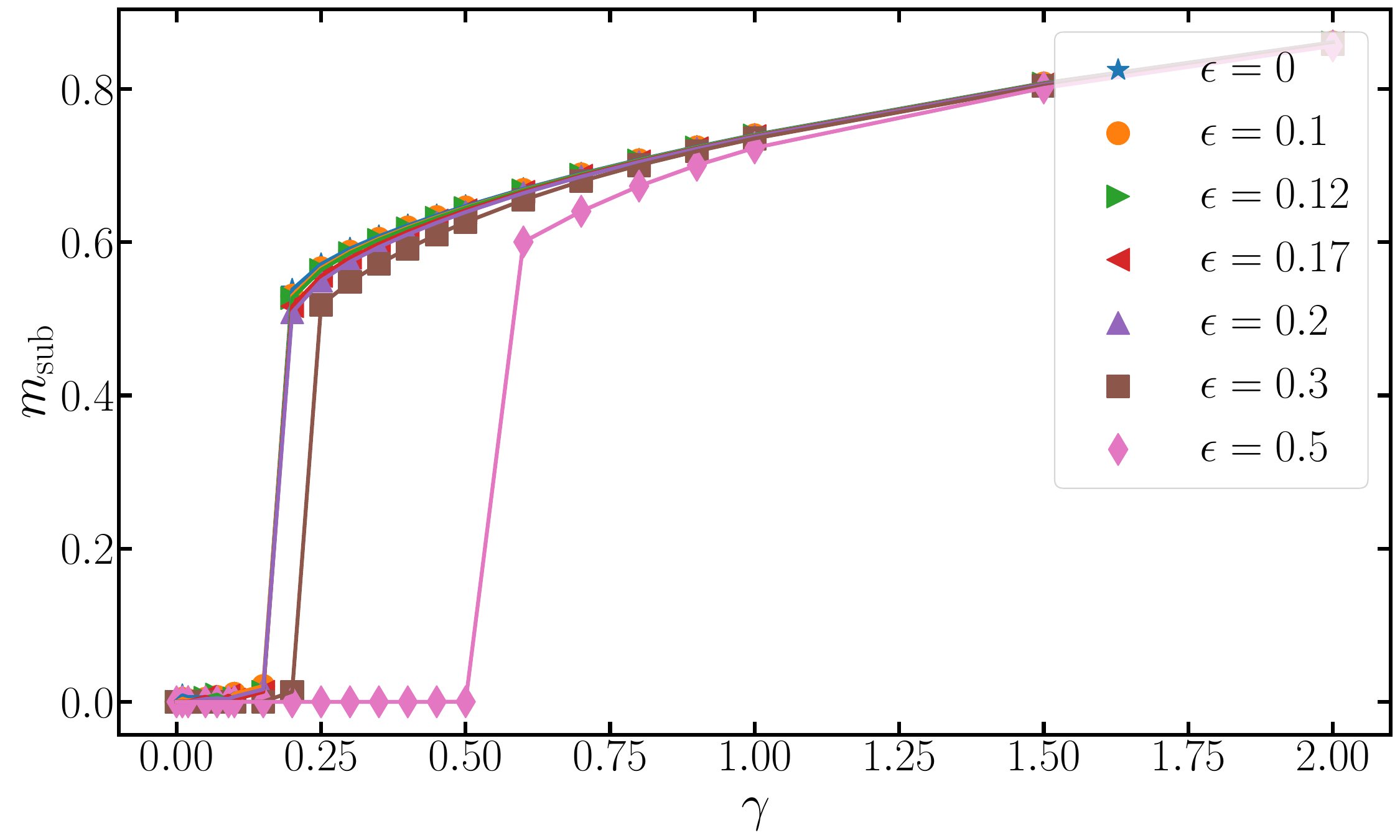}
  \caption{Subharmonic response $\msub$ as a function of the measurement strength $\gamma$ along the most-probable trajectory for pulse imperfections $\epsilon=0$--$0.5$. N\'eel state, $L=14$, $J=2$, $g=0.5$, $\alpha=1$.}
  \label{fig:eps_cuts}
\end{figure}

\section{Trajectory-averaged results for the unbounded protocol ($w\to\infty$)}
\label{app:unbounded}

In this Appendix, we present the numerical results for the unbounded stochastic protocol ($w\to\infty$) discussed in Sec.~\ref{sec:res_unbounded}. Figures~\ref{fig:unbounded}(a)--\ref{fig:unbounded}(c) show the trajectory-averaged entanglement entropy $\avtraj{\Sent(nT)}$ and Figs.~\ref{fig:unbounded}(d)--\ref{fig:unbounded}(f) show the trajectory-averaged stroboscopic magnetization $M(nT)$, for $\gamma=0$, 0.1, and 2 and the N\'eel initial state.

We first discuss the entanglement. Without measurements ($\gamma=0$), the entropy saturates at a value that grows with $L$, as expected for volume-law scaling. For weak monitoring ($\gamma=0.1$), the entropy saturates faster and at lower values, but it still increases clearly with $L$; the system thus remains in the volume-law phase. For strong monitoring ($\gamma=2$), the entropy saturates at a small value, $\overline{\avtraj{\Sent}}\approx0.25$, which is independent of $L$. This is the area-law phase. The unbounded protocol therefore shows a measurement-induced transition from volume-law to area-law entanglement as $\gamma$ increases.

The magnetization behaves very differently. For all three values of $\gamma$, $M(nT)$ decays to zero within a few Floquet periods and shows no persistent period-doubled oscillation. At $\gamma=0$, the remaining fluctuations are large, since the dynamics is unitary and no trajectory average is taken; they decrease with increasing $L$. For $\gamma>0$, the fluctuations around zero are much smaller. Consequently, the subharmonic component $|\overline{(-1)^nM}|$ vanishes for all values of $\gamma$ considered, consistent with the heating of the ensemble-averaged state discussed in the main text. Figure~\ref{fig:unbounded} thus confirms that the area-law phase at $\gamma=2$ does not come with DTC order: in the unbounded protocol, suppressed entanglement and period doubling are not connected.

\begin{figure}[t]
  \centering
  \includegraphics[width=\columnwidth]{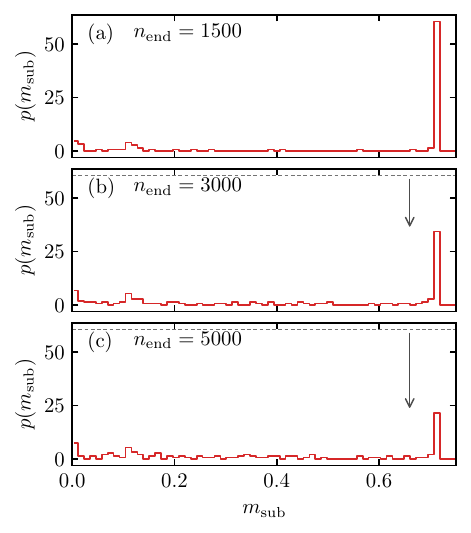}
  \caption{Histogram of the late-time subharmonic response $m^{(r)}_{\mathrm{sub}}$ over trajectories at $w=0.4$ for late-time windows ending at (a) $n_{\mathrm{end}}=1500$, (b) $n_{\mathrm{end}}=3000$, and (c) $n_{\mathrm{end}}=5000$, with $n_0=1200$. The dashed lines in (b) and (c) mark the height of the time-crystalline peak in (a); arrows indicate its decrease. Parameters as in Fig.~\ref{fig:hist}.}
  \label{fig:hist_time}
\end{figure}

\section{Subharmonic response for different pulse imperfections ($w=0$)}
\label{app:epsilon}

Figure~\ref{fig:eps_cuts} shows $\msub$ as a function of $\gamma$ for several values of $\epsilon$ along the most-probable trajectory for the N\'eel state with $L=14$. These are cuts through Fig.~\ref{fig:phase}(a). For $\epsilon\le0.2$, the curves are almost identical, with onset near $\gamma\approx0.2$. For larger $\epsilon$, the onset shifts to stronger measurement, reaching $\gamma\approx0.6$ at $\epsilon=0.5$. Above the onset, all curves approach the same value, confirming that the response is rigid against pulse imperfections.

\begin{figure}[t]
  \centering
  \includegraphics[width=0.92\columnwidth]{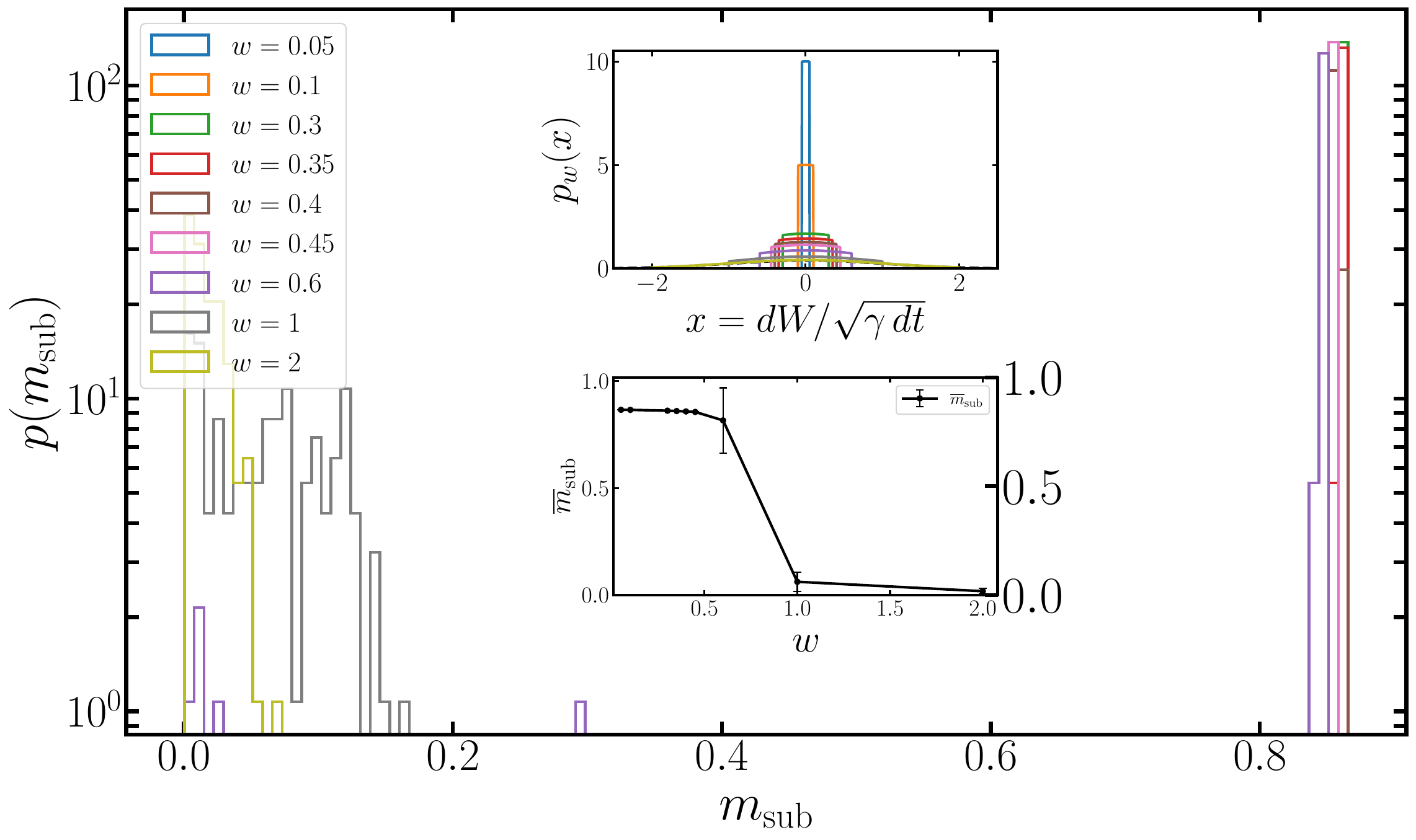}
  \caption{Distribution $p(\msub)$ of the late-time subharmonic response for different postselection widths $w=0.05$--$2$ for $\alpha=2$. N\'eel state, $L=14$, $\epsilon=0.15$, $\gamma=1$. Upper inset: bounded-Gaussian noise distributions $p_w(x)$, together with the standard Gaussian. Lower inset: mean response $\avtraj{\msub}$ versus $w$, with error bars indicating the trajectory-to-trajectory spread.}
  \label{fig:alpha}
\end{figure}

\section{Time dependence of the subharmonic-response histogram}
\label{app:hist_time}

Figure~\ref{fig:hist_time} shows the histogram of $m^{(r)}_{\mathrm{sub}}$ at $w=0.4$ for three late-time windows, ending at $n_{\mathrm{end}}=1500$, 3000, and 5000. For all $n_{\mathrm{end}}$, the histogram has a sharp peak at $\msub\approx0.72$ and a broad component at small $\msub$. As $n_{\mathrm{end}}$ increases, the weight of the time-crystalline peak decreases steadily: between $n_{\mathrm{end}}=1500$ and 5000, its height drops to roughly one third of its initial value, while the weight at small and intermediate $\msub$ grows correspondingly. The peak position, in contrast, stays fixed. Trajectories therefore leave the time-crystalline class one by one, rather than losing their amplitude gradually. This explains the decrease of $\avtraj{\msub}$ with $n_{\mathrm{end}}$ at intermediate $w$ (Fig.~\ref{fig:nend}).

\section{Robustness to the interaction range}
\label{app:alpha}

To test whether our conclusions depend on the range of the interactions, we repeat the bounded-protocol analysis for a shorter-ranged Ising coupling, $\alpha=2$ in Eq.~\eqref{eq:H}. Figure~\ref{fig:alpha} shows the full distribution $p(\msub)$ for a sequence of postselection widths. The qualitative picture is the same as for $\alpha=1$ (Sec.~\ref{sec:res_bounded}). For narrow windows, all trajectories sit in a single sharp peak, here at $\msub\approx0.85$; for wide windows, the weight moves to $\msub\lesssim0.2$ and the period doubling is lost. In between, the distribution is nearly bimodal, reflecting the coexistence of time-crystalline and non-time-crystalline trajectories. The mean response $\avtraj{\msub}$ (lower inset) remains at this value up to $w\approx0.6$ and then drops to nearly zero. Compared with $\alpha=1$, where $\avtraj{\msub}$ already drops sharply between $w=0.3$ and $0.45$ (Fig.~\ref{fig:nend}), increasing $\alpha$ thus increases the range of noise windows $w$ over which the period-doubled response survives, so that shorter-ranged interactions make the DTC more tolerant to measurement noise. The stabilization of period doubling by bounded measurement noise is therefore not specific to a particular interaction range.


\bibliography{apssamp}

\begin{thebibliography}{79}%
\makeatletter
\providecommand \@ifxundefined [1]{%
 \@ifx{#1\undefined}
}%
\providecommand \@ifnum [1]{%
 \ifnum #1\expandafter \@firstoftwo
 \else \expandafter \@secondoftwo
 \fi
}%
\providecommand \@ifx [1]{%
 \ifx #1\expandafter \@firstoftwo
 \else \expandafter \@secondoftwo
 \fi
}%
\providecommand \natexlab [1]{#1}%
\providecommand \enquote  [1]{``#1''}%
\providecommand \bibnamefont  [1]{#1}%
\providecommand \bibfnamefont [1]{#1}%
\providecommand \citenamefont [1]{#1}%
\providecommand \href@noop [0]{\@secondoftwo}%
\providecommand \href [0]{\begingroup \@sanitize@url \@href}%
\providecommand \@href[1]{\@@startlink{#1}\@@href}%
\providecommand \@@href[1]{\endgroup#1\@@endlink}%
\providecommand \@sanitize@url [0]{\catcode `\\12\catcode `\$12\catcode
  `\&12\catcode `\#12\catcode `\^12\catcode `\_12\catcode `\%12\relax}%
\providecommand \@@startlink[1]{}%
\providecommand \@@endlink[0]{}%
\providecommand \url  [0]{\begingroup\@sanitize@url \@url }%
\providecommand \@url [1]{\endgroup\@href {#1}{\urlprefix }}%
\providecommand \urlprefix  [0]{URL }%
\providecommand \Eprint [0]{\href }%
\providecommand \doibase [0]{https://doi.org/}%
\providecommand \selectlanguage [0]{\@gobble}%
\providecommand \bibinfo  [0]{\@secondoftwo}%
\providecommand \bibfield  [0]{\@secondoftwo}%
\providecommand \translation [1]{[#1]}%
\providecommand \BibitemOpen [0]{}%
\providecommand \bibitemStop [0]{}%
\providecommand \bibitemNoStop [0]{.\EOS\space}%
\providecommand \EOS [0]{\spacefactor3000\relax}%
\providecommand \BibitemShut  [1]{\csname bibitem#1\endcsname}%
\let\auto@bib@innerbib\@empty
\bibitem [{\citenamefont {Shapere}\ and\ \citenamefont
  {Wilczek}(2012)}]{DTC_th_1}%
  \BibitemOpen
  \bibfield  {author} {\bibinfo {author} {\bibfnamefont {A.}~\bibnamefont
  {Shapere}}\ and\ \bibinfo {author} {\bibfnamefont {F.}~\bibnamefont
  {Wilczek}},\ }\bibfield  {title} {\bibinfo {title} {Classical time
  crystals},\ }\href {https://doi.org/10.1103/PhysRevLett.109.160402}
  {\bibfield  {journal} {\bibinfo  {journal} {Phys. Rev. Lett.}\ }\textbf
  {\bibinfo {volume} {109}},\ \bibinfo {pages} {160402} (\bibinfo {year}
  {2012})}\BibitemShut {NoStop}%
\bibitem [{\citenamefont {Wilczek}(2012)}]{DTC_th_2}%
  \BibitemOpen
  \bibfield  {author} {\bibinfo {author} {\bibfnamefont {F.}~\bibnamefont
  {Wilczek}},\ }\bibfield  {title} {\bibinfo {title} {Quantum time crystals},\
  }\href {https://doi.org/10.1103/PhysRevLett.109.160401} {\bibfield  {journal}
  {\bibinfo  {journal} {Phys. Rev. Lett.}\ }\textbf {\bibinfo {volume} {109}},\
  \bibinfo {pages} {160401} (\bibinfo {year} {2012})}\BibitemShut {NoStop}%
\bibitem [{\citenamefont {Wilczek}(2013)}]{DTC_th_3}%
  \BibitemOpen
  \bibfield  {author} {\bibinfo {author} {\bibfnamefont {F.}~\bibnamefont
  {Wilczek}},\ }\bibfield  {title} {\bibinfo {title} {Superfluidity and
  space-time translation symmetry breaking},\ }\href
  {https://doi.org/10.1103/PhysRevLett.111.250402} {\bibfield  {journal}
  {\bibinfo  {journal} {Phys. Rev. Lett.}\ }\textbf {\bibinfo {volume} {111}},\
  \bibinfo {pages} {250402} (\bibinfo {year} {2013})}\BibitemShut {NoStop}%
\bibitem [{\citenamefont {Bruno}(2013)}]{DTC_th_4}%
  \BibitemOpen
  \bibfield  {author} {\bibinfo {author} {\bibfnamefont {P.}~\bibnamefont
  {Bruno}},\ }\bibfield  {title} {\bibinfo {title} {Comment on ``{Quantum Time
  Crystals}''},\ }\href {https://doi.org/10.1103/PhysRevLett.110.118901}
  {\bibfield  {journal} {\bibinfo  {journal} {Phys. Rev. Lett.}\ }\textbf
  {\bibinfo {volume} {110}},\ \bibinfo {pages} {118901} (\bibinfo {year}
  {2013})}\BibitemShut {NoStop}%
\bibitem [{\citenamefont {Watanabe}\ and\ \citenamefont
  {Oshikawa}(2015)}]{DTC_th_5}%
  \BibitemOpen
  \bibfield  {author} {\bibinfo {author} {\bibfnamefont {H.}~\bibnamefont
  {Watanabe}}\ and\ \bibinfo {author} {\bibfnamefont {M.}~\bibnamefont
  {Oshikawa}},\ }\bibfield  {title} {\bibinfo {title} {Absence of quantum time
  crystals},\ }\href {https://doi.org/10.1103/PhysRevLett.114.251603}
  {\bibfield  {journal} {\bibinfo  {journal} {Phys. Rev. Lett.}\ }\textbf
  {\bibinfo {volume} {114}},\ \bibinfo {pages} {251603} (\bibinfo {year}
  {2015})}\BibitemShut {NoStop}%
\bibitem [{\citenamefont {Sacha}(2015)}]{DTC_th_6}%
  \BibitemOpen
  \bibfield  {author} {\bibinfo {author} {\bibfnamefont {K.}~\bibnamefont
  {Sacha}},\ }\bibfield  {title} {\bibinfo {title} {Modeling spontaneous
  breaking of time-translation symmetry},\ }\href
  {https://doi.org/10.1103/PhysRevA.91.033617} {\bibfield  {journal} {\bibinfo
  {journal} {Phys. Rev. A}\ }\textbf {\bibinfo {volume} {91}},\ \bibinfo
  {pages} {033617} (\bibinfo {year} {2015})}\BibitemShut {NoStop}%
\bibitem [{\citenamefont {Chandran}\ and\ \citenamefont
  {Sondhi}(2016)}]{DTC_th_7}%
  \BibitemOpen
  \bibfield  {author} {\bibinfo {author} {\bibfnamefont {A.}~\bibnamefont
  {Chandran}}\ and\ \bibinfo {author} {\bibfnamefont {S.~L.}\ \bibnamefont
  {Sondhi}},\ }\bibfield  {title} {\bibinfo {title} {Interaction-stabilized
  steady states in the driven {$O(N)$} model},\ }\href
  {https://doi.org/10.1103/PhysRevB.93.174305} {\bibfield  {journal} {\bibinfo
  {journal} {Phys. Rev. B}\ }\textbf {\bibinfo {volume} {93}},\ \bibinfo
  {pages} {174305} (\bibinfo {year} {2016})}\BibitemShut {NoStop}%
\bibitem [{\citenamefont {Khemani}\ \emph {et~al.}(2016)\citenamefont
  {Khemani}, \citenamefont {Lazarides}, \citenamefont {Moessner},\ and\
  \citenamefont {Sondhi}}]{DTC_th_8}%
  \BibitemOpen
  \bibfield  {author} {\bibinfo {author} {\bibfnamefont {V.}~\bibnamefont
  {Khemani}}, \bibinfo {author} {\bibfnamefont {A.}~\bibnamefont {Lazarides}},
  \bibinfo {author} {\bibfnamefont {R.}~\bibnamefont {Moessner}},\ and\
  \bibinfo {author} {\bibfnamefont {S.~L.}\ \bibnamefont {Sondhi}},\ }\bibfield
   {title} {\bibinfo {title} {Phase structure of driven quantum systems},\
  }\href {https://doi.org/10.1103/PhysRevLett.116.250401} {\bibfield  {journal}
  {\bibinfo  {journal} {Phys. Rev. Lett.}\ }\textbf {\bibinfo {volume} {116}},\
  \bibinfo {pages} {250401} (\bibinfo {year} {2016})}\BibitemShut {NoStop}%
\bibitem [{\citenamefont {Else}\ \emph {et~al.}(2016)\citenamefont {Else},
  \citenamefont {Bauer},\ and\ \citenamefont {Nayak}}]{DTC_th_9}%
  \BibitemOpen
  \bibfield  {author} {\bibinfo {author} {\bibfnamefont {D.~V.}\ \bibnamefont
  {Else}}, \bibinfo {author} {\bibfnamefont {B.}~\bibnamefont {Bauer}},\ and\
  \bibinfo {author} {\bibfnamefont {C.}~\bibnamefont {Nayak}},\ }\bibfield
  {title} {\bibinfo {title} {Floquet time crystals},\ }\href
  {https://doi.org/10.1103/PhysRevLett.117.090402} {\bibfield  {journal}
  {\bibinfo  {journal} {Phys. Rev. Lett.}\ }\textbf {\bibinfo {volume} {117}},\
  \bibinfo {pages} {090402} (\bibinfo {year} {2016})}\BibitemShut {NoStop}%
\bibitem [{\citenamefont {von Keyserlingk}\ \emph {et~al.}(2016)\citenamefont
  {von Keyserlingk}, \citenamefont {Khemani},\ and\ \citenamefont
  {Sondhi}}]{DTC_th_10}%
  \BibitemOpen
  \bibfield  {author} {\bibinfo {author} {\bibfnamefont {C.~W.}\ \bibnamefont
  {von Keyserlingk}}, \bibinfo {author} {\bibfnamefont {V.}~\bibnamefont
  {Khemani}},\ and\ \bibinfo {author} {\bibfnamefont {S.~L.}\ \bibnamefont
  {Sondhi}},\ }\bibfield  {title} {\bibinfo {title} {Absolute stability and
  spatiotemporal long-range order in {Floquet} systems},\ }\href
  {https://doi.org/10.1103/PhysRevB.94.085112} {\bibfield  {journal} {\bibinfo
  {journal} {Phys. Rev. B}\ }\textbf {\bibinfo {volume} {94}},\ \bibinfo
  {pages} {085112} (\bibinfo {year} {2016})}\BibitemShut {NoStop}%
\bibitem [{\citenamefont {Yao}\ \emph {et~al.}(2017)\citenamefont {Yao},
  \citenamefont {Potter}, \citenamefont {Potirniche},\ and\ \citenamefont
  {Vishwanath}}]{DTC_th_11}%
  \BibitemOpen
  \bibfield  {author} {\bibinfo {author} {\bibfnamefont {N.~Y.}\ \bibnamefont
  {Yao}}, \bibinfo {author} {\bibfnamefont {A.~C.}\ \bibnamefont {Potter}},
  \bibinfo {author} {\bibfnamefont {I.-D.}\ \bibnamefont {Potirniche}},\ and\
  \bibinfo {author} {\bibfnamefont {A.}~\bibnamefont {Vishwanath}},\ }\bibfield
   {title} {\bibinfo {title} {Discrete time crystals: Rigidity, criticality,
  and realizations},\ }\href {https://doi.org/10.1103/PhysRevLett.118.030401}
  {\bibfield  {journal} {\bibinfo  {journal} {Phys. Rev. Lett.}\ }\textbf
  {\bibinfo {volume} {118}},\ \bibinfo {pages} {030401} (\bibinfo {year}
  {2017})}\BibitemShut {NoStop}%
\bibitem [{\citenamefont {Khemani}\ \emph {et~al.}(2021)\citenamefont
  {Khemani}, \citenamefont {Moessner},\ and\ \citenamefont
  {Sondhi}}]{khemani2021comment}%
  \BibitemOpen
  \bibfield  {author} {\bibinfo {author} {\bibfnamefont {V.}~\bibnamefont
  {Khemani}}, \bibinfo {author} {\bibfnamefont {R.}~\bibnamefont {Moessner}},\
  and\ \bibinfo {author} {\bibfnamefont {S.}~\bibnamefont {Sondhi}},\
  }\bibfield  {title} {\bibinfo {title} {A comment on ``{Discrete time
  crystals: Rigidity, criticality, and realizations}''},\ }\href@noop {}
  {\bibfield  {journal} {\bibinfo  {journal} {arXiv:2109.00551}\ } (\bibinfo
  {year} {2021})}\BibitemShut {NoStop}%
\bibitem [{\citenamefont {Zaletel}\ \emph {et~al.}(2023)\citenamefont
  {Zaletel}, \citenamefont {Lukin}, \citenamefont {Monroe}, \citenamefont
  {Nayak}, \citenamefont {Wilczek},\ and\ \citenamefont {Yao}}]{Zaletel2023}%
  \BibitemOpen
  \bibfield  {author} {\bibinfo {author} {\bibfnamefont {M.~P.}\ \bibnamefont
  {Zaletel}}, \bibinfo {author} {\bibfnamefont {M.}~\bibnamefont {Lukin}},
  \bibinfo {author} {\bibfnamefont {C.}~\bibnamefont {Monroe}}, \bibinfo
  {author} {\bibfnamefont {C.}~\bibnamefont {Nayak}}, \bibinfo {author}
  {\bibfnamefont {F.}~\bibnamefont {Wilczek}},\ and\ \bibinfo {author}
  {\bibfnamefont {N.~Y.}\ \bibnamefont {Yao}},\ }\bibfield  {title} {\bibinfo
  {title} {Colloquium: Quantum and classical discrete time crystals},\
  }\href@noop {} {\bibfield  {journal} {\bibinfo  {journal} {Rev. Mod. Phys.}\
  }\textbf {\bibinfo {volume} {95}},\ \bibinfo {pages} {031001} (\bibinfo
  {year} {2023})}\BibitemShut {NoStop}%
\bibitem [{\citenamefont {Agrawal}\ \emph {et~al.}(2026)\citenamefont
  {Agrawal}, \citenamefont {Vasista}, \citenamefont {Nag},\ and\ \citenamefont
  {Choudhury}}]{agrawal2026quantum}%
  \BibitemOpen
  \bibfield  {author} {\bibinfo {author} {\bibfnamefont {A.}~\bibnamefont
  {Agrawal}}, \bibinfo {author} {\bibfnamefont {A.~A.}\ \bibnamefont
  {Vasista}}, \bibinfo {author} {\bibfnamefont {T.}~\bibnamefont {Nag}},\ and\
  \bibinfo {author} {\bibfnamefont {S.}~\bibnamefont {Choudhury}},\ }\bibfield
  {title} {\bibinfo {title} {Quantum state revivals, time quasicrystals, and
  entanglement steering in a fibonacci-driven central-spin model},\ }\href@noop
  {} {\bibfield  {journal} {\bibinfo  {journal} {arXiv preprint
  arXiv:2610.35555}\ } (\bibinfo {year} {2026})}\BibitemShut {NoStop}%
\bibitem [{\citenamefont {Zhang}\ \emph {et~al.}(2017)\citenamefont {Zhang},
  \citenamefont {Hess}, \citenamefont {Kyprianidis}, \citenamefont {Becker},
  \citenamefont {Lee}, \citenamefont {Smith}, \citenamefont {Pagano},
  \citenamefont {Potirniche}, \citenamefont {Potter}, \citenamefont
  {Vishwanath}, \citenamefont {Yao},\ and\ \citenamefont
  {Monroe}}]{exp_nature_1}%
  \BibitemOpen
  \bibfield  {author} {\bibinfo {author} {\bibfnamefont {J.}~\bibnamefont
  {Zhang}}, \bibinfo {author} {\bibfnamefont {P.~W.}\ \bibnamefont {Hess}},
  \bibinfo {author} {\bibfnamefont {A.}~\bibnamefont {Kyprianidis}}, \bibinfo
  {author} {\bibfnamefont {P.}~\bibnamefont {Becker}}, \bibinfo {author}
  {\bibfnamefont {A.}~\bibnamefont {Lee}}, \bibinfo {author} {\bibfnamefont
  {J.}~\bibnamefont {Smith}}, \bibinfo {author} {\bibfnamefont
  {G.}~\bibnamefont {Pagano}}, \bibinfo {author} {\bibfnamefont {I.-D.}\
  \bibnamefont {Potirniche}}, \bibinfo {author} {\bibfnamefont {A.~C.}\
  \bibnamefont {Potter}}, \bibinfo {author} {\bibfnamefont {A.}~\bibnamefont
  {Vishwanath}}, \bibinfo {author} {\bibfnamefont {N.~Y.}\ \bibnamefont
  {Yao}},\ and\ \bibinfo {author} {\bibfnamefont {C.}~\bibnamefont {Monroe}},\
  }\bibfield  {title} {\bibinfo {title} {Observation of a discrete time
  crystal},\ }\href {https://doi.org/10.1038/nature21413} {\bibfield  {journal}
  {\bibinfo  {journal} {Nature}\ }\textbf {\bibinfo {volume} {543}},\ \bibinfo
  {pages} {217} (\bibinfo {year} {2017})}\BibitemShut {NoStop}%
\bibitem [{\citenamefont {Choi}\ \emph {et~al.}(2017)\citenamefont {Choi},
  \citenamefont {Choi}, \citenamefont {Landig}, \citenamefont {Kucsko},
  \citenamefont {Zhou}, \citenamefont {Isoya}, \citenamefont {Jelezko},
  \citenamefont {Onoda}, \citenamefont {Sumiya}, \citenamefont {Khemani},
  \citenamefont {von Keyserlingk}, \citenamefont {Yao}, \citenamefont
  {Demler},\ and\ \citenamefont {Lukin}}]{exp_nature_2}%
  \BibitemOpen
  \bibfield  {author} {\bibinfo {author} {\bibfnamefont {S.}~\bibnamefont
  {Choi}}, \bibinfo {author} {\bibfnamefont {J.}~\bibnamefont {Choi}}, \bibinfo
  {author} {\bibfnamefont {R.}~\bibnamefont {Landig}}, \bibinfo {author}
  {\bibfnamefont {G.}~\bibnamefont {Kucsko}}, \bibinfo {author} {\bibfnamefont
  {H.}~\bibnamefont {Zhou}}, \bibinfo {author} {\bibfnamefont {J.}~\bibnamefont
  {Isoya}}, \bibinfo {author} {\bibfnamefont {F.}~\bibnamefont {Jelezko}},
  \bibinfo {author} {\bibfnamefont {S.}~\bibnamefont {Onoda}}, \bibinfo
  {author} {\bibfnamefont {H.}~\bibnamefont {Sumiya}}, \bibinfo {author}
  {\bibfnamefont {V.}~\bibnamefont {Khemani}}, \bibinfo {author} {\bibfnamefont
  {C.}~\bibnamefont {von Keyserlingk}}, \bibinfo {author} {\bibfnamefont
  {N.~Y.}\ \bibnamefont {Yao}}, \bibinfo {author} {\bibfnamefont
  {E.}~\bibnamefont {Demler}},\ and\ \bibinfo {author} {\bibfnamefont {M.~D.}\
  \bibnamefont {Lukin}},\ }\bibfield  {title} {\bibinfo {title} {Observation of
  discrete time-crystalline order in a disordered dipolar many-body system},\
  }\href {https://doi.org/10.1038/nature21426} {\bibfield  {journal} {\bibinfo
  {journal} {Nature}\ }\textbf {\bibinfo {volume} {543}},\ \bibinfo {pages}
  {221} (\bibinfo {year} {2017})}\BibitemShut {NoStop}%
\bibitem [{\citenamefont {Huang}\ \emph {et~al.}(2018)\citenamefont {Huang},
  \citenamefont {Wu},\ and\ \citenamefont {Liu}}]{exp_3}%
  \BibitemOpen
  \bibfield  {author} {\bibinfo {author} {\bibfnamefont {B.}~\bibnamefont
  {Huang}}, \bibinfo {author} {\bibfnamefont {Y.-H.}\ \bibnamefont {Wu}},\ and\
  \bibinfo {author} {\bibfnamefont {W.~V.}\ \bibnamefont {Liu}},\ }\bibfield
  {title} {\bibinfo {title} {Clean {Floquet} time crystals: Models and
  realizations in cold atoms},\ }\href
  {https://doi.org/10.1103/PhysRevLett.120.110603} {\bibfield  {journal}
  {\bibinfo  {journal} {Phys. Rev. Lett.}\ }\textbf {\bibinfo {volume} {120}},\
  \bibinfo {pages} {110603} (\bibinfo {year} {2018})}\BibitemShut {NoStop}%
\bibitem [{\citenamefont {W\"uster}\ \emph {et~al.}(2012)\citenamefont
  {W\"uster}, \citenamefont {D{\k{a}}browska-W\"uster},\ and\ \citenamefont
  {Davis}}]{exp_4}%
  \BibitemOpen
  \bibfield  {author} {\bibinfo {author} {\bibfnamefont {S.}~\bibnamefont
  {W\"uster}}, \bibinfo {author} {\bibfnamefont {B.~J.}\ \bibnamefont
  {D{\k{a}}browska-W\"uster}},\ and\ \bibinfo {author} {\bibfnamefont {M.~J.}\
  \bibnamefont {Davis}},\ }\bibfield  {title} {\bibinfo {title} {Macroscopic
  quantum self-trapping in dynamical tunneling},\ }\href
  {https://doi.org/10.1103/PhysRevLett.109.080401} {\bibfield  {journal}
  {\bibinfo  {journal} {Phys. Rev. Lett.}\ }\textbf {\bibinfo {volume} {109}},\
  \bibinfo {pages} {080401} (\bibinfo {year} {2012})}\BibitemShut {NoStop}%
\bibitem [{\citenamefont {Pizzi}\ \emph {et~al.}(2021)\citenamefont {Pizzi},
  \citenamefont {Knolle},\ and\ \citenamefont {Nunnenkamp}}]{exp_5}%
  \BibitemOpen
  \bibfield  {author} {\bibinfo {author} {\bibfnamefont {A.}~\bibnamefont
  {Pizzi}}, \bibinfo {author} {\bibfnamefont {J.}~\bibnamefont {Knolle}},\ and\
  \bibinfo {author} {\bibfnamefont {A.}~\bibnamefont {Nunnenkamp}},\ }\bibfield
   {title} {\bibinfo {title} {Higher-order and fractional discrete time
  crystals in clean long-range interacting systems},\ }\href
  {https://doi.org/10.1038/s41467-021-22822-3} {\bibfield  {journal} {\bibinfo
  {journal} {Nat. Commun.}\ }\textbf {\bibinfo {volume} {12}},\ \bibinfo
  {pages} {2341} (\bibinfo {year} {2021})}\BibitemShut {NoStop}%
\bibitem [{\citenamefont {Russomanno}\ \emph {et~al.}(2017)\citenamefont
  {Russomanno}, \citenamefont {Iemini}, \citenamefont {Dalmonte},\ and\
  \citenamefont {Fazio}}]{exp_th_1}%
  \BibitemOpen
  \bibfield  {author} {\bibinfo {author} {\bibfnamefont {A.}~\bibnamefont
  {Russomanno}}, \bibinfo {author} {\bibfnamefont {F.}~\bibnamefont {Iemini}},
  \bibinfo {author} {\bibfnamefont {M.}~\bibnamefont {Dalmonte}},\ and\
  \bibinfo {author} {\bibfnamefont {R.}~\bibnamefont {Fazio}},\ }\bibfield
  {title} {\bibinfo {title} {Floquet time crystal in the {Lipkin-Meshkov-Glick}
  model},\ }\href {https://doi.org/10.1103/PhysRevB.95.214307} {\bibfield
  {journal} {\bibinfo  {journal} {Phys. Rev. B}\ }\textbf {\bibinfo {volume}
  {95}},\ \bibinfo {pages} {214307} (\bibinfo {year} {2017})}\BibitemShut
  {NoStop}%
\bibitem [{\citenamefont {Giergiel}\ \emph {et~al.}(2019)\citenamefont
  {Giergiel}, \citenamefont {Kuro\ifmmode~\acute{s}\else \'{s}\fi{}},\ and\
  \citenamefont {Sacha}}]{exp_th_2}%
  \BibitemOpen
  \bibfield  {author} {\bibinfo {author} {\bibfnamefont {K.}~\bibnamefont
  {Giergiel}}, \bibinfo {author} {\bibfnamefont {A.}~\bibnamefont
  {Kuro\ifmmode~\acute{s}\else \'{s}\fi{}}},\ and\ \bibinfo {author}
  {\bibfnamefont {K.}~\bibnamefont {Sacha}},\ }\bibfield  {title} {\bibinfo
  {title} {Discrete time quasicrystals},\ }\href
  {https://doi.org/10.1103/PhysRevB.99.220303} {\bibfield  {journal} {\bibinfo
  {journal} {Phys. Rev. B}\ }\textbf {\bibinfo {volume} {99}},\ \bibinfo
  {pages} {220303(R)} (\bibinfo {year} {2019})}\BibitemShut {NoStop}%
\bibitem [{\citenamefont {Matus}\ and\ \citenamefont {Sacha}(2019)}]{exp_th_3}%
  \BibitemOpen
  \bibfield  {author} {\bibinfo {author} {\bibfnamefont {P.}~\bibnamefont
  {Matus}}\ and\ \bibinfo {author} {\bibfnamefont {K.}~\bibnamefont {Sacha}},\
  }\bibfield  {title} {\bibinfo {title} {Fractional time crystals},\ }\href
  {https://doi.org/10.1103/PhysRevA.99.033626} {\bibfield  {journal} {\bibinfo
  {journal} {Phys. Rev. A}\ }\textbf {\bibinfo {volume} {99}},\ \bibinfo
  {pages} {033626} (\bibinfo {year} {2019})}\BibitemShut {NoStop}%
\bibitem [{\citenamefont {Yousefjani}\ \emph
  {et~al.}(2025{\natexlab{a}})\citenamefont {Yousefjani}, \citenamefont
  {Sacha},\ and\ \citenamefont {Bayat}}]{exp_th_4}%
  \BibitemOpen
  \bibfield  {author} {\bibinfo {author} {\bibfnamefont {R.}~\bibnamefont
  {Yousefjani}}, \bibinfo {author} {\bibfnamefont {K.}~\bibnamefont {Sacha}},\
  and\ \bibinfo {author} {\bibfnamefont {A.}~\bibnamefont {Bayat}},\ }\bibfield
   {title} {\bibinfo {title} {Discrete time crystal phase as a resource for
  quantum-enhanced sensing},\ }\href
  {https://doi.org/10.1103/PhysRevB.111.125159} {\bibfield  {journal} {\bibinfo
   {journal} {Phys. Rev. B}\ }\textbf {\bibinfo {volume} {111}},\ \bibinfo
  {pages} {125159} (\bibinfo {year} {2025}{\natexlab{a}})}\BibitemShut
  {NoStop}%
\bibitem [{\citenamefont {Liu}\ \emph {et~al.}(2023)\citenamefont {Liu},
  \citenamefont {Zhang}, \citenamefont {Hsieh}, \citenamefont {Zhang},\ and\
  \citenamefont {Yao}}]{exp_th_5}%
  \BibitemOpen
  \bibfield  {author} {\bibinfo {author} {\bibfnamefont {S.}~\bibnamefont
  {Liu}}, \bibinfo {author} {\bibfnamefont {S.-X.}\ \bibnamefont {Zhang}},
  \bibinfo {author} {\bibfnamefont {C.-Y.}\ \bibnamefont {Hsieh}}, \bibinfo
  {author} {\bibfnamefont {S.}~\bibnamefont {Zhang}},\ and\ \bibinfo {author}
  {\bibfnamefont {H.}~\bibnamefont {Yao}},\ }\bibfield  {title} {\bibinfo
  {title} {Discrete time crystal enabled by {Stark} many-body localization},\
  }\href {https://doi.org/10.1103/PhysRevLett.130.120403} {\bibfield  {journal}
  {\bibinfo  {journal} {Phys. Rev. Lett.}\ }\textbf {\bibinfo {volume} {130}},\
  \bibinfo {pages} {120403} (\bibinfo {year} {2023})}\BibitemShut {NoStop}%
\bibitem [{\citenamefont {Kshetrimayum}\ \emph {et~al.}(2020)\citenamefont
  {Kshetrimayum}, \citenamefont {Eisert},\ and\ \citenamefont
  {Kennes}}]{exp_th_6}%
  \BibitemOpen
  \bibfield  {author} {\bibinfo {author} {\bibfnamefont {A.}~\bibnamefont
  {Kshetrimayum}}, \bibinfo {author} {\bibfnamefont {J.}~\bibnamefont
  {Eisert}},\ and\ \bibinfo {author} {\bibfnamefont {D.~M.}\ \bibnamefont
  {Kennes}},\ }\bibfield  {title} {\bibinfo {title} {Stark time crystals:
  Symmetry breaking in space and time},\ }\href
  {https://doi.org/10.1103/PhysRevB.102.195116} {\bibfield  {journal} {\bibinfo
   {journal} {Phys. Rev. B}\ }\textbf {\bibinfo {volume} {102}},\ \bibinfo
  {pages} {195116} (\bibinfo {year} {2020})}\BibitemShut {NoStop}%
\bibitem [{\citenamefont {Collura}\ \emph {et~al.}(2022)\citenamefont
  {Collura}, \citenamefont {De~Luca}, \citenamefont {Rossini},\ and\
  \citenamefont {Lerose}}]{exp_th_7}%
  \BibitemOpen
  \bibfield  {author} {\bibinfo {author} {\bibfnamefont {M.}~\bibnamefont
  {Collura}}, \bibinfo {author} {\bibfnamefont {A.}~\bibnamefont {De~Luca}},
  \bibinfo {author} {\bibfnamefont {D.}~\bibnamefont {Rossini}},\ and\ \bibinfo
  {author} {\bibfnamefont {A.}~\bibnamefont {Lerose}},\ }\bibfield  {title}
  {\bibinfo {title} {Discrete time-crystalline response stabilized by
  domain-wall confinement},\ }\href
  {https://doi.org/10.1103/PhysRevX.12.031037} {\bibfield  {journal} {\bibinfo
  {journal} {Phys. Rev. X}\ }\textbf {\bibinfo {volume} {12}},\ \bibinfo
  {pages} {031037} (\bibinfo {year} {2022})}\BibitemShut {NoStop}%
\bibitem [{\citenamefont {Maskara}\ \emph {et~al.}(2021)\citenamefont
  {Maskara}, \citenamefont {Michailidis}, \citenamefont {Ho}, \citenamefont
  {Bluvstein}, \citenamefont {Choi}, \citenamefont {Lukin},\ and\ \citenamefont
  {Serbyn}}]{ext_th_8}%
  \BibitemOpen
  \bibfield  {author} {\bibinfo {author} {\bibfnamefont {N.}~\bibnamefont
  {Maskara}}, \bibinfo {author} {\bibfnamefont {A.~A.}\ \bibnamefont
  {Michailidis}}, \bibinfo {author} {\bibfnamefont {W.~W.}\ \bibnamefont {Ho}},
  \bibinfo {author} {\bibfnamefont {D.}~\bibnamefont {Bluvstein}}, \bibinfo
  {author} {\bibfnamefont {S.}~\bibnamefont {Choi}}, \bibinfo {author}
  {\bibfnamefont {M.~D.}\ \bibnamefont {Lukin}},\ and\ \bibinfo {author}
  {\bibfnamefont {M.}~\bibnamefont {Serbyn}},\ }\bibfield  {title} {\bibinfo
  {title} {Discrete time-crystalline order enabled by quantum many-body scars:
  Entanglement steering via periodic driving},\ }\href
  {https://doi.org/10.1103/PhysRevLett.127.090602} {\bibfield  {journal}
  {\bibinfo  {journal} {Phys. Rev. Lett.}\ }\textbf {\bibinfo {volume} {127}},\
  \bibinfo {pages} {090602} (\bibinfo {year} {2021})}\BibitemShut {NoStop}%
\bibitem [{\citenamefont {Deng}\ and\ \citenamefont {Yang}(2023)}]{exp_th_9}%
  \BibitemOpen
  \bibfield  {author} {\bibinfo {author} {\bibfnamefont {W.}~\bibnamefont
  {Deng}}\ and\ \bibinfo {author} {\bibfnamefont {Z.-C.}\ \bibnamefont
  {Yang}},\ }\bibfield  {title} {\bibinfo {title} {Using models with static
  quantum many-body scars to generate time-crystalline behavior under periodic
  driving},\ }\href {https://doi.org/10.1103/PhysRevB.108.205129} {\bibfield
  {journal} {\bibinfo  {journal} {Phys. Rev. B}\ }\textbf {\bibinfo {volume}
  {108}},\ \bibinfo {pages} {205129} (\bibinfo {year} {2023})}\BibitemShut
  {NoStop}%
\bibitem [{\citenamefont {Bull}\ \emph {et~al.}(2022)\citenamefont {Bull},
  \citenamefont {Hallam}, \citenamefont {Papi\ifmmode~\acute{c}\else
  \'{c}\fi{}},\ and\ \citenamefont {Martin}}]{exp_th_10}%
  \BibitemOpen
  \bibfield  {author} {\bibinfo {author} {\bibfnamefont {K.}~\bibnamefont
  {Bull}}, \bibinfo {author} {\bibfnamefont {A.}~\bibnamefont {Hallam}},
  \bibinfo {author} {\bibfnamefont {Z.}~\bibnamefont
  {Papi\ifmmode~\acute{c}\else \'{c}\fi{}}},\ and\ \bibinfo {author}
  {\bibfnamefont {I.}~\bibnamefont {Martin}},\ }\bibfield  {title} {\bibinfo
  {title} {Tuning between continuous time crystals and many-body scars in
  long-range {$XYZ$} spin chains},\ }\href
  {https://doi.org/10.1103/PhysRevLett.129.140602} {\bibfield  {journal}
  {\bibinfo  {journal} {Phys. Rev. Lett.}\ }\textbf {\bibinfo {volume} {129}},\
  \bibinfo {pages} {140602} (\bibinfo {year} {2022})}\BibitemShut {NoStop}%
\bibitem [{\citenamefont {Huang}\ \emph {et~al.}(2022)\citenamefont {Huang},
  \citenamefont {Leung}, \citenamefont {Stamper-Kurn},\ and\ \citenamefont
  {Liu}}]{exp_th_11}%
  \BibitemOpen
  \bibfield  {author} {\bibinfo {author} {\bibfnamefont {B.}~\bibnamefont
  {Huang}}, \bibinfo {author} {\bibfnamefont {T.-H.}\ \bibnamefont {Leung}},
  \bibinfo {author} {\bibfnamefont {D.~M.}\ \bibnamefont {Stamper-Kurn}},\ and\
  \bibinfo {author} {\bibfnamefont {W.~V.}\ \bibnamefont {Liu}},\ }\bibfield
  {title} {\bibinfo {title} {Discrete time crystals enforced by {Floquet-Bloch}
  scars},\ }\href {https://doi.org/10.1103/PhysRevLett.129.133001} {\bibfield
  {journal} {\bibinfo  {journal} {Phys. Rev. Lett.}\ }\textbf {\bibinfo
  {volume} {129}},\ \bibinfo {pages} {133001} (\bibinfo {year}
  {2022})}\BibitemShut {NoStop}%
\bibitem [{\citenamefont {Huang}(2023)}]{exp_th_12}%
  \BibitemOpen
  \bibfield  {author} {\bibinfo {author} {\bibfnamefont {B.}~\bibnamefont
  {Huang}},\ }\bibfield  {title} {\bibinfo {title} {Analytical theory of cat
  scars with discrete time-crystalline dynamics in {Floquet} systems},\ }\href
  {https://doi.org/10.1103/PhysRevB.108.104309} {\bibfield  {journal} {\bibinfo
   {journal} {Phys. Rev. B}\ }\textbf {\bibinfo {volume} {108}},\ \bibinfo
  {pages} {104309} (\bibinfo {year} {2023})}\BibitemShut {NoStop}%
\bibitem [{\citenamefont {Oka}\ and\ \citenamefont
  {Kitamura}(2019)}]{oka2019floquet}%
  \BibitemOpen
  \bibfield  {author} {\bibinfo {author} {\bibfnamefont {T.}~\bibnamefont
  {Oka}}\ and\ \bibinfo {author} {\bibfnamefont {S.}~\bibnamefont {Kitamura}},\
  }\bibfield  {title} {\bibinfo {title} {Floquet engineering of quantum
  materials},\ }\href@noop {} {\bibfield  {journal} {\bibinfo  {journal}
  {Annual Review of Condensed Matter Physics}\ }\textbf {\bibinfo {volume}
  {10}},\ \bibinfo {pages} {387} (\bibinfo {year} {2019})}\BibitemShut
  {NoStop}%
\bibitem [{\citenamefont {Mondal}\ \emph {et~al.}(2026)\citenamefont {Mondal},
  \citenamefont {Ghosh}, \citenamefont {Saha},\ and\ \citenamefont
  {Nag}}]{mondal2026floquet}%
  \BibitemOpen
  \bibfield  {author} {\bibinfo {author} {\bibfnamefont {D.}~\bibnamefont
  {Mondal}}, \bibinfo {author} {\bibfnamefont {A.~K.}\ \bibnamefont {Ghosh}},
  \bibinfo {author} {\bibfnamefont {A.}~\bibnamefont {Saha}},\ and\ \bibinfo
  {author} {\bibfnamefont {T.}~\bibnamefont {Nag}},\ }\bibfield  {title}
  {\bibinfo {title} {Floquet first-and higher-order topological
  superconductors: A review},\ }\href@noop {} {\bibfield  {journal} {\bibinfo
  {journal} {Journal of Physics: Condensed Matter}\ }\textbf {\bibinfo {volume}
  {38}},\ \bibinfo {pages} {433001} (\bibinfo {year} {2026})}\BibitemShut
  {NoStop}%
\bibitem [{\citenamefont {Ghosh}\ \emph {et~al.}(2024)\citenamefont {Ghosh},
  \citenamefont {Nag},\ and\ \citenamefont {Saha}}]{ghosh2024generation}%
  \BibitemOpen
  \bibfield  {author} {\bibinfo {author} {\bibfnamefont {A.~K.}\ \bibnamefont
  {Ghosh}}, \bibinfo {author} {\bibfnamefont {T.}~\bibnamefont {Nag}},\ and\
  \bibinfo {author} {\bibfnamefont {A.}~\bibnamefont {Saha}},\ }\bibfield
  {title} {\bibinfo {title} {Generation of higher-order topological insulators
  using periodic driving},\ }\href@noop {} {\bibfield  {journal} {\bibinfo
  {journal} {Journal of Physics: Condensed Matter}\ }\textbf {\bibinfo {volume}
  {36}},\ \bibinfo {pages} {093001} (\bibinfo {year} {2024})}\BibitemShut
  {NoStop}%
\bibitem [{\citenamefont {Pal}\ \emph {et~al.}(2025)\citenamefont {Pal},
  \citenamefont {Mondal}, \citenamefont {Nag},\ and\ \citenamefont
  {Saha}}]{prnx-47mk}%
  \BibitemOpen
  \bibfield  {author} {\bibinfo {author} {\bibfnamefont {A.}~\bibnamefont
  {Pal}}, \bibinfo {author} {\bibfnamefont {D.}~\bibnamefont {Mondal}},
  \bibinfo {author} {\bibfnamefont {T.}~\bibnamefont {Nag}},\ and\ \bibinfo
  {author} {\bibfnamefont {A.}~\bibnamefont {Saha}},\ }\bibfield  {title}
  {\bibinfo {title} {Josephson current signature of floquet majorana and
  topological accidental zero modes in altermagnet heterostructures},\ }\href
  {https://doi.org/10.1103/prnx-47mk} {\bibfield  {journal} {\bibinfo
  {journal} {Phys. Rev. B}\ }\textbf {\bibinfo {volume} {112}},\ \bibinfo
  {pages} {L201408} (\bibinfo {year} {2025})}\BibitemShut {NoStop}%
\bibitem [{\citenamefont {Martin}\ \emph {et~al.}(2017)\citenamefont {Martin},
  \citenamefont {Refael},\ and\ \citenamefont {Halperin}}]{PhysRevX.7.041008}%
  \BibitemOpen
  \bibfield  {author} {\bibinfo {author} {\bibfnamefont {I.}~\bibnamefont
  {Martin}}, \bibinfo {author} {\bibfnamefont {G.}~\bibnamefont {Refael}},\
  and\ \bibinfo {author} {\bibfnamefont {B.}~\bibnamefont {Halperin}},\
  }\bibfield  {title} {\bibinfo {title} {Topological frequency conversion in
  strongly driven quantum systems},\ }\href
  {https://doi.org/10.1103/PhysRevX.7.041008} {\bibfield  {journal} {\bibinfo
  {journal} {Phys. Rev. X}\ }\textbf {\bibinfo {volume} {7}},\ \bibinfo {pages}
  {041008} (\bibinfo {year} {2017})}\BibitemShut {NoStop}%
\bibitem [{\citenamefont {Kaushik}\ \emph {et~al.}(2026)\citenamefont
  {Kaushik}, \citenamefont {Choudhury},\ and\ \citenamefont {Nag}}]{prg5-3srn}%
  \BibitemOpen
  \bibfield  {author} {\bibinfo {author} {\bibfnamefont {V.}~\bibnamefont
  {Kaushik}}, \bibinfo {author} {\bibfnamefont {S.}~\bibnamefont {Choudhury}},\
  and\ \bibinfo {author} {\bibfnamefont {T.}~\bibnamefont {Nag}},\ }\bibfield
  {title} {\bibinfo {title} {Topological energy pumping in a quasiperiodically
  driven four-level system},\ }\href {https://doi.org/10.1103/prg5-3srn}
  {\bibfield  {journal} {\bibinfo  {journal} {Phys. Rev. B}\ }\textbf {\bibinfo
  {volume} {113}},\ \bibinfo {pages} {184312} (\bibinfo {year}
  {2026})}\BibitemShut {NoStop}%
\bibitem [{\citenamefont {Mori}(2023)}]{mori2023floquet}%
  \BibitemOpen
  \bibfield  {author} {\bibinfo {author} {\bibfnamefont {T.}~\bibnamefont
  {Mori}},\ }\bibfield  {title} {\bibinfo {title} {Floquet states in open
  quantum systems},\ }\href@noop {} {\bibfield  {journal} {\bibinfo  {journal}
  {Annual Review of Condensed Matter Physics}\ }\textbf {\bibinfo {volume}
  {14}},\ \bibinfo {pages} {35} (\bibinfo {year} {2023})}\BibitemShut {NoStop}%
\bibitem [{\citenamefont {Chen}\ \emph {et~al.}(2024)\citenamefont {Chen},
  \citenamefont {Hu}, \citenamefont {Zhang}, \citenamefont {Kurniawan},
  \citenamefont {Shao}, \citenamefont {Chen}, \citenamefont {Prem},\ and\
  \citenamefont {Dai}}]{PhysRevB.109.184309}%
  \BibitemOpen
  \bibfield  {author} {\bibinfo {author} {\bibfnamefont {H.}~\bibnamefont
  {Chen}}, \bibinfo {author} {\bibfnamefont {Y.-M.}\ \bibnamefont {Hu}},
  \bibinfo {author} {\bibfnamefont {W.}~\bibnamefont {Zhang}}, \bibinfo
  {author} {\bibfnamefont {M.~A.}\ \bibnamefont {Kurniawan}}, \bibinfo {author}
  {\bibfnamefont {Y.}~\bibnamefont {Shao}}, \bibinfo {author} {\bibfnamefont
  {X.}~\bibnamefont {Chen}}, \bibinfo {author} {\bibfnamefont {A.}~\bibnamefont
  {Prem}},\ and\ \bibinfo {author} {\bibfnamefont {X.}~\bibnamefont {Dai}},\
  }\bibfield  {title} {\bibinfo {title} {Periodically driven open quantum
  systems: Spectral properties and nonequilibrium steady states},\ }\href
  {https://doi.org/10.1103/PhysRevB.109.184309} {\bibfield  {journal} {\bibinfo
   {journal} {Phys. Rev. B}\ }\textbf {\bibinfo {volume} {109}},\ \bibinfo
  {pages} {184309} (\bibinfo {year} {2024})}\BibitemShut {NoStop}%
\bibitem [{\citenamefont {Gupt}\ \emph {et~al.}(2022)\citenamefont {Gupt},
  \citenamefont {Bhattacharyya}, \citenamefont {Das}, \citenamefont {Datta},
  \citenamefont {Mukherjee},\ and\ \citenamefont
  {Ghosh}}]{PhysRevE.106.024110}%
  \BibitemOpen
  \bibfield  {author} {\bibinfo {author} {\bibfnamefont {N.}~\bibnamefont
  {Gupt}}, \bibinfo {author} {\bibfnamefont {S.}~\bibnamefont {Bhattacharyya}},
  \bibinfo {author} {\bibfnamefont {B.}~\bibnamefont {Das}}, \bibinfo {author}
  {\bibfnamefont {S.}~\bibnamefont {Datta}}, \bibinfo {author} {\bibfnamefont
  {V.}~\bibnamefont {Mukherjee}},\ and\ \bibinfo {author} {\bibfnamefont
  {A.}~\bibnamefont {Ghosh}},\ }\bibfield  {title} {\bibinfo {title} {Floquet
  quantum thermal transistor},\ }\href
  {https://doi.org/10.1103/PhysRevE.106.024110} {\bibfield  {journal} {\bibinfo
   {journal} {Phys. Rev. E}\ }\textbf {\bibinfo {volume} {106}},\ \bibinfo
  {pages} {024110} (\bibinfo {year} {2022})}\BibitemShut {NoStop}%
\bibitem [{\citenamefont {Cuetara}\ \emph {et~al.}(2015)\citenamefont
  {Cuetara}, \citenamefont {Engel},\ and\ \citenamefont
  {Esposito}}]{cuetara2015stochastic}%
  \BibitemOpen
  \bibfield  {author} {\bibinfo {author} {\bibfnamefont {G.~B.}\ \bibnamefont
  {Cuetara}}, \bibinfo {author} {\bibfnamefont {A.}~\bibnamefont {Engel}},\
  and\ \bibinfo {author} {\bibfnamefont {M.}~\bibnamefont {Esposito}},\
  }\bibfield  {title} {\bibinfo {title} {Stochastic thermodynamics of rapidly
  driven systems},\ }\href@noop {} {\bibfield  {journal} {\bibinfo  {journal}
  {New journal of physics}\ }\textbf {\bibinfo {volume} {17}},\ \bibinfo
  {pages} {055002} (\bibinfo {year} {2015})}\BibitemShut {NoStop}%
\bibitem [{\citenamefont {Basko}\ \emph {et~al.}(2006)\citenamefont {Basko},
  \citenamefont {Aleiner},\ and\ \citenamefont {Altshuler}}]{Basko2006}%
  \BibitemOpen
  \bibfield  {author} {\bibinfo {author} {\bibfnamefont {D.~M.}\ \bibnamefont
  {Basko}}, \bibinfo {author} {\bibfnamefont {I.~L.}\ \bibnamefont {Aleiner}},\
  and\ \bibinfo {author} {\bibfnamefont {B.~L.}\ \bibnamefont {Altshuler}},\
  }\bibfield  {title} {\bibinfo {title} {Metal--insulator transition in a
  weakly interacting many-electron system with localized single-particle
  states},\ }\href {https://doi.org/10.1016/j.aop.2005.11.014} {\bibfield
  {journal} {\bibinfo  {journal} {Ann. Phys.}\ }\textbf {\bibinfo {volume}
  {321}},\ \bibinfo {pages} {1126} (\bibinfo {year} {2006})}\BibitemShut
  {NoStop}%
\bibitem [{\citenamefont {Pal}\ and\ \citenamefont {Huse}(2010)}]{Pal2010}%
  \BibitemOpen
  \bibfield  {author} {\bibinfo {author} {\bibfnamefont {A.}~\bibnamefont
  {Pal}}\ and\ \bibinfo {author} {\bibfnamefont {D.~A.}\ \bibnamefont {Huse}},\
  }\bibfield  {title} {\bibinfo {title} {Many-body localization phase
  transition},\ }\href {https://doi.org/10.1103/PhysRevB.82.174411} {\bibfield
  {journal} {\bibinfo  {journal} {Phys. Rev. B}\ }\textbf {\bibinfo {volume}
  {82}},\ \bibinfo {pages} {174411} (\bibinfo {year} {2010})}\BibitemShut
  {NoStop}%
\bibitem [{\citenamefont {Nandkishore}\ and\ \citenamefont
  {Huse}(2015)}]{Nandkishore2015}%
  \BibitemOpen
  \bibfield  {author} {\bibinfo {author} {\bibfnamefont {R.}~\bibnamefont
  {Nandkishore}}\ and\ \bibinfo {author} {\bibfnamefont {D.~A.}\ \bibnamefont
  {Huse}},\ }\bibfield  {title} {\bibinfo {title} {Many-body localization and
  thermalization in quantum statistical mechanics},\ }\href
  {https://doi.org/10.1146/annurev-conmatphys-031214-014726} {\bibfield
  {journal} {\bibinfo  {journal} {Annu. Rev. Condens. Matter Phys.}\ }\textbf
  {\bibinfo {volume} {6}},\ \bibinfo {pages} {15} (\bibinfo {year}
  {2015})}\BibitemShut {NoStop}%
\bibitem [{\citenamefont {Abanin}\ \emph {et~al.}(2019)\citenamefont {Abanin},
  \citenamefont {Altman}, \citenamefont {Bloch},\ and\ \citenamefont
  {Serbyn}}]{Abanin2019}%
  \BibitemOpen
  \bibfield  {author} {\bibinfo {author} {\bibfnamefont {D.~A.}\ \bibnamefont
  {Abanin}}, \bibinfo {author} {\bibfnamefont {E.}~\bibnamefont {Altman}},
  \bibinfo {author} {\bibfnamefont {I.}~\bibnamefont {Bloch}},\ and\ \bibinfo
  {author} {\bibfnamefont {M.}~\bibnamefont {Serbyn}},\ }\bibfield  {title}
  {\bibinfo {title} {Colloquium: Many-body localization, thermalization, and
  entanglement},\ }\href {https://doi.org/10.1103/RevModPhys.91.021001}
  {\bibfield  {journal} {\bibinfo  {journal} {Rev. Mod. Phys.}\ }\textbf
  {\bibinfo {volume} {91}},\ \bibinfo {pages} {021001} (\bibinfo {year}
  {2019})}\BibitemShut {NoStop}%
\bibitem [{\citenamefont {Modak}\ and\ \citenamefont
  {Nag}(2020)}]{PhysRevE.101.052108}%
  \BibitemOpen
  \bibfield  {author} {\bibinfo {author} {\bibfnamefont {R.}~\bibnamefont
  {Modak}}\ and\ \bibinfo {author} {\bibfnamefont {T.}~\bibnamefont {Nag}},\
  }\bibfield  {title} {\bibinfo {title} {Many-body localization in a long-range
  model: Real-space renormalization-group study},\ }\href
  {https://doi.org/10.1103/PhysRevE.101.052108} {\bibfield  {journal} {\bibinfo
   {journal} {Phys. Rev. E}\ }\textbf {\bibinfo {volume} {101}},\ \bibinfo
  {pages} {052108} (\bibinfo {year} {2020})}\BibitemShut {NoStop}%
\bibitem [{\citenamefont {Modak}\ and\ \citenamefont
  {Mukerjee}(2015)}]{PhysRevLett.115.230401}%
  \BibitemOpen
  \bibfield  {author} {\bibinfo {author} {\bibfnamefont {R.}~\bibnamefont
  {Modak}}\ and\ \bibinfo {author} {\bibfnamefont {S.}~\bibnamefont
  {Mukerjee}},\ }\bibfield  {title} {\bibinfo {title} {Many-body localization
  in the presence of a single-particle mobility edge},\ }\href
  {https://doi.org/10.1103/PhysRevLett.115.230401} {\bibfield  {journal}
  {\bibinfo  {journal} {Phys. Rev. Lett.}\ }\textbf {\bibinfo {volume} {115}},\
  \bibinfo {pages} {230401} (\bibinfo {year} {2015})}\BibitemShut {NoStop}%
\bibitem [{\citenamefont {Ponte}\ \emph {et~al.}(2015)\citenamefont {Ponte},
  \citenamefont {Papi{\'c}}, \citenamefont {Huveneers},\ and\ \citenamefont
  {Abanin}}]{Ponte2015}%
  \BibitemOpen
  \bibfield  {author} {\bibinfo {author} {\bibfnamefont {P.}~\bibnamefont
  {Ponte}}, \bibinfo {author} {\bibfnamefont {Z.}~\bibnamefont {Papi{\'c}}},
  \bibinfo {author} {\bibfnamefont {F.}~\bibnamefont {Huveneers}},\ and\
  \bibinfo {author} {\bibfnamefont {D.~A.}\ \bibnamefont {Abanin}},\ }\bibfield
   {title} {\bibinfo {title} {Many-body localization in periodically driven
  systems},\ }\href {https://doi.org/10.1103/PhysRevLett.114.140401} {\bibfield
   {journal} {\bibinfo  {journal} {Phys. Rev. Lett.}\ }\textbf {\bibinfo
  {volume} {114}},\ \bibinfo {pages} {140401} (\bibinfo {year}
  {2015})}\BibitemShut {NoStop}%
\bibitem [{\citenamefont {Lazarides}\ \emph {et~al.}(2015)\citenamefont
  {Lazarides}, \citenamefont {Das},\ and\ \citenamefont
  {Moessner}}]{Lazarides2015}%
  \BibitemOpen
  \bibfield  {author} {\bibinfo {author} {\bibfnamefont {A.}~\bibnamefont
  {Lazarides}}, \bibinfo {author} {\bibfnamefont {A.}~\bibnamefont {Das}},\
  and\ \bibinfo {author} {\bibfnamefont {R.}~\bibnamefont {Moessner}},\
  }\bibfield  {title} {\bibinfo {title} {Fate of many-body localization under
  periodic driving},\ }\href {https://doi.org/10.1103/PhysRevLett.115.030402}
  {\bibfield  {journal} {\bibinfo  {journal} {Phys. Rev. Lett.}\ }\textbf
  {\bibinfo {volume} {115}},\ \bibinfo {pages} {030402} (\bibinfo {year}
  {2015})}\BibitemShut {NoStop}%
\bibitem [{\citenamefont {Else}\ \emph {et~al.}(2017)\citenamefont {Else},
  \citenamefont {Bauer},\ and\ \citenamefont {Nayak}}]{Else2017}%
  \BibitemOpen
  \bibfield  {author} {\bibinfo {author} {\bibfnamefont {D.~V.}\ \bibnamefont
  {Else}}, \bibinfo {author} {\bibfnamefont {B.}~\bibnamefont {Bauer}},\ and\
  \bibinfo {author} {\bibfnamefont {C.}~\bibnamefont {Nayak}},\ }\bibfield
  {title} {\bibinfo {title} {Prethermal phases of matter protected by
  time-translation symmetry},\ }\href@noop {} {\bibfield  {journal} {\bibinfo
  {journal} {Phys. Rev. X}\ }\textbf {\bibinfo {volume} {7}},\ \bibinfo {pages}
  {011026} (\bibinfo {year} {2017})}\BibitemShut {NoStop}%
\bibitem [{\citenamefont {Machado}\ \emph {et~al.}(2020)\citenamefont
  {Machado}, \citenamefont {Else}, \citenamefont {Kahanamoku-Meyer},
  \citenamefont {Nayak},\ and\ \citenamefont {Yao}}]{Machado2020}%
  \BibitemOpen
  \bibfield  {author} {\bibinfo {author} {\bibfnamefont {F.}~\bibnamefont
  {Machado}}, \bibinfo {author} {\bibfnamefont {D.~V.}\ \bibnamefont {Else}},
  \bibinfo {author} {\bibfnamefont {G.~D.}\ \bibnamefont {Kahanamoku-Meyer}},
  \bibinfo {author} {\bibfnamefont {C.}~\bibnamefont {Nayak}},\ and\ \bibinfo
  {author} {\bibfnamefont {N.~Y.}\ \bibnamefont {Yao}},\ }\bibfield  {title}
  {\bibinfo {title} {Long-range prethermal phases of nonequilibrium matter},\
  }\href@noop {} {\bibfield  {journal} {\bibinfo  {journal} {Phys. Rev. X}\
  }\textbf {\bibinfo {volume} {10}},\ \bibinfo {pages} {011043} (\bibinfo
  {year} {2020})}\BibitemShut {NoStop}%
\bibitem [{\citenamefont {Lazarides}\ \emph {et~al.}(2020)\citenamefont
  {Lazarides}, \citenamefont {Roy}, \citenamefont {Piazza},\ and\ \citenamefont
  {Moessner}}]{open_DTC_4}%
  \BibitemOpen
  \bibfield  {author} {\bibinfo {author} {\bibfnamefont {A.}~\bibnamefont
  {Lazarides}}, \bibinfo {author} {\bibfnamefont {S.}~\bibnamefont {Roy}},
  \bibinfo {author} {\bibfnamefont {F.}~\bibnamefont {Piazza}},\ and\ \bibinfo
  {author} {\bibfnamefont {R.}~\bibnamefont {Moessner}},\ }\bibfield  {title}
  {\bibinfo {title} {Time crystallinity in dissipative {Floquet} systems},\
  }\href {https://doi.org/10.1103/PhysRevResearch.2.022002} {\bibfield
  {journal} {\bibinfo  {journal} {Phys. Rev. Res.}\ }\textbf {\bibinfo {volume}
  {2}},\ \bibinfo {pages} {022002(R)} (\bibinfo {year} {2020})}\BibitemShut
  {NoStop}%
\bibitem [{\citenamefont {Kyprianidis}\ \emph {et~al.}(2021)\citenamefont
  {Kyprianidis}, \citenamefont {Machado}, \citenamefont {Morong}, \citenamefont
  {Becker}, \citenamefont {Collins}, \citenamefont {Else}, \citenamefont
  {Feng}, \citenamefont {Hess}, \citenamefont {Nayak}, \citenamefont {Pagano},
  \citenamefont {Yao},\ and\ \citenamefont
  {Monroe}}]{doi:10.1126/science.abg8102}%
  \BibitemOpen
  \bibfield  {author} {\bibinfo {author} {\bibfnamefont {A.}~\bibnamefont
  {Kyprianidis}}, \bibinfo {author} {\bibfnamefont {F.}~\bibnamefont
  {Machado}}, \bibinfo {author} {\bibfnamefont {W.}~\bibnamefont {Morong}},
  \bibinfo {author} {\bibfnamefont {P.}~\bibnamefont {Becker}}, \bibinfo
  {author} {\bibfnamefont {K.~S.}\ \bibnamefont {Collins}}, \bibinfo {author}
  {\bibfnamefont {D.~V.}\ \bibnamefont {Else}}, \bibinfo {author}
  {\bibfnamefont {L.}~\bibnamefont {Feng}}, \bibinfo {author} {\bibfnamefont
  {P.~W.}\ \bibnamefont {Hess}}, \bibinfo {author} {\bibfnamefont
  {C.}~\bibnamefont {Nayak}}, \bibinfo {author} {\bibfnamefont
  {G.}~\bibnamefont {Pagano}}, \bibinfo {author} {\bibfnamefont {N.~Y.}\
  \bibnamefont {Yao}},\ and\ \bibinfo {author} {\bibfnamefont {C.}~\bibnamefont
  {Monroe}},\ }\bibfield  {title} {\bibinfo {title} {Observation of a
  prethermal discrete time crystal},\ }\href
  {https://doi.org/10.1126/science.abg8102} {\bibfield  {journal} {\bibinfo
  {journal} {Science}\ }\textbf {\bibinfo {volume} {372}},\ \bibinfo {pages}
  {1192} (\bibinfo {year} {2021})}\BibitemShut {NoStop}%
\bibitem [{\citenamefont {Lyu}\ \emph {et~al.}(2020)\citenamefont {Lyu},
  \citenamefont {Choudhury}, \citenamefont {Lv}, \citenamefont {Yan},\ and\
  \citenamefont {Zhou}}]{prethermal_2}%
  \BibitemOpen
  \bibfield  {author} {\bibinfo {author} {\bibfnamefont {C.}~\bibnamefont
  {Lyu}}, \bibinfo {author} {\bibfnamefont {S.}~\bibnamefont {Choudhury}},
  \bibinfo {author} {\bibfnamefont {C.}~\bibnamefont {Lv}}, \bibinfo {author}
  {\bibfnamefont {Y.}~\bibnamefont {Yan}},\ and\ \bibinfo {author}
  {\bibfnamefont {Q.}~\bibnamefont {Zhou}},\ }\bibfield  {title} {\bibinfo
  {title} {Eternal discrete time crystal beating the {Heisenberg} limit},\
  }\href {https://doi.org/10.1103/PhysRevResearch.2.033070} {\bibfield
  {journal} {\bibinfo  {journal} {Phys. Rev. Res.}\ }\textbf {\bibinfo {volume}
  {2}},\ \bibinfo {pages} {033070} (\bibinfo {year} {2020})}\BibitemShut
  {NoStop}%
\bibitem [{\citenamefont {Bothra}\ and\ \citenamefont
  {Garg}(2026)}]{Bothra2025}%
  \BibitemOpen
  \bibfield  {author} {\bibinfo {author} {\bibfnamefont {A.}~\bibnamefont
  {Bothra}}\ and\ \bibinfo {author} {\bibfnamefont {A.}~\bibnamefont {Garg}},\
  }\bibfield  {title} {\bibinfo {title} {Discrete time crystal in periodically
  driven quantum {Sherrington-Kirkpatrick} model},\ }\href
  {https://doi.org/10.1103/qw9c-zl7h} {\bibfield  {journal} {\bibinfo
  {journal} {Phys. Rev. B}\ }\textbf {\bibinfo {volume} {114}},\ \bibinfo
  {pages} {194202} (\bibinfo {year} {2026})}\BibitemShut {NoStop}%
\bibitem [{\citenamefont {Lazarides}\ and\ \citenamefont
  {Moessner}(2017)}]{open_DTC_destroy_1}%
  \BibitemOpen
  \bibfield  {author} {\bibinfo {author} {\bibfnamefont {A.}~\bibnamefont
  {Lazarides}}\ and\ \bibinfo {author} {\bibfnamefont {R.}~\bibnamefont
  {Moessner}},\ }\bibfield  {title} {\bibinfo {title} {Fate of a discrete time
  crystal in an open system},\ }\href
  {https://doi.org/10.1103/PhysRevB.95.195135} {\bibfield  {journal} {\bibinfo
  {journal} {Phys. Rev. B}\ }\textbf {\bibinfo {volume} {95}},\ \bibinfo
  {pages} {195135} (\bibinfo {year} {2017})}\BibitemShut {NoStop}%
\bibitem [{\citenamefont {Riera-Campeny}\ \emph {et~al.}(2020)\citenamefont
  {Riera-Campeny}, \citenamefont {Moreno-Cardoner},\ and\ \citenamefont
  {Sanpera}}]{open_DTC_destroy_2}%
  \BibitemOpen
  \bibfield  {author} {\bibinfo {author} {\bibfnamefont {A.}~\bibnamefont
  {Riera-Campeny}}, \bibinfo {author} {\bibfnamefont {M.}~\bibnamefont
  {Moreno-Cardoner}},\ and\ \bibinfo {author} {\bibfnamefont {A.}~\bibnamefont
  {Sanpera}},\ }\bibfield  {title} {\bibinfo {title} {Time crystallinity in
  open quantum systems},\ }\href {https://doi.org/10.22331/q-2020-05-25-270}
  {\bibfield  {journal} {\bibinfo  {journal} {{Quantum}}\ }\textbf {\bibinfo
  {volume} {4}},\ \bibinfo {pages} {270} (\bibinfo {year} {2020})}\BibitemShut
  {NoStop}%
\bibitem [{\citenamefont {Yousefjani}\ \emph
  {et~al.}(2025{\natexlab{b}})\citenamefont {Yousefjani}, \citenamefont
  {Carollo}, \citenamefont {Sacha}, \citenamefont {Al-Kuwari},\ and\
  \citenamefont {Bayat}}]{open_DTC_1}%
  \BibitemOpen
  \bibfield  {author} {\bibinfo {author} {\bibfnamefont {R.}~\bibnamefont
  {Yousefjani}}, \bibinfo {author} {\bibfnamefont {A.}~\bibnamefont {Carollo}},
  \bibinfo {author} {\bibfnamefont {K.}~\bibnamefont {Sacha}}, \bibinfo
  {author} {\bibfnamefont {S.}~\bibnamefont {Al-Kuwari}},\ and\ \bibinfo
  {author} {\bibfnamefont {A.}~\bibnamefont {Bayat}},\ }\bibfield  {title}
  {\bibinfo {title} {{Non-Hermitian} discrete time crystals},\ }\href
  {https://doi.org/10.1103/PhysRevB.111.165117} {\bibfield  {journal} {\bibinfo
   {journal} {Phys. Rev. B}\ }\textbf {\bibinfo {volume} {111}},\ \bibinfo
  {pages} {165117} (\bibinfo {year} {2025}{\natexlab{b}})}\BibitemShut
  {NoStop}%
\bibitem [{\citenamefont {Gong}\ \emph {et~al.}(2018)\citenamefont {Gong},
  \citenamefont {Ashida}, \citenamefont {Kawabata}, \citenamefont {Takasan},
  \citenamefont {Higashikawa},\ and\ \citenamefont {Ueda}}]{open_DTC_2}%
  \BibitemOpen
  \bibfield  {author} {\bibinfo {author} {\bibfnamefont {Z.}~\bibnamefont
  {Gong}}, \bibinfo {author} {\bibfnamefont {Y.}~\bibnamefont {Ashida}},
  \bibinfo {author} {\bibfnamefont {K.}~\bibnamefont {Kawabata}}, \bibinfo
  {author} {\bibfnamefont {K.}~\bibnamefont {Takasan}}, \bibinfo {author}
  {\bibfnamefont {S.}~\bibnamefont {Higashikawa}},\ and\ \bibinfo {author}
  {\bibfnamefont {M.}~\bibnamefont {Ueda}},\ }\bibfield  {title} {\bibinfo
  {title} {Topological phases of {Non-Hermitian} systems},\ }\href
  {https://doi.org/10.1103/PhysRevX.8.031079} {\bibfield  {journal} {\bibinfo
  {journal} {Phys. Rev. X}\ }\textbf {\bibinfo {volume} {8}},\ \bibinfo {pages}
  {031079} (\bibinfo {year} {2018})}\BibitemShut {NoStop}%
\bibitem [{\citenamefont {Gambetta}\ \emph {et~al.}(2019)\citenamefont
  {Gambetta}, \citenamefont {Carollo}, \citenamefont {Marcuzzi}, \citenamefont
  {Garrahan},\ and\ \citenamefont {Lesanovsky}}]{open_DTC_3}%
  \BibitemOpen
  \bibfield  {author} {\bibinfo {author} {\bibfnamefont {F.~M.}\ \bibnamefont
  {Gambetta}}, \bibinfo {author} {\bibfnamefont {F.}~\bibnamefont {Carollo}},
  \bibinfo {author} {\bibfnamefont {M.}~\bibnamefont {Marcuzzi}}, \bibinfo
  {author} {\bibfnamefont {J.~P.}\ \bibnamefont {Garrahan}},\ and\ \bibinfo
  {author} {\bibfnamefont {I.}~\bibnamefont {Lesanovsky}},\ }\bibfield  {title}
  {\bibinfo {title} {Discrete time crystals in the absence of manifest
  symmetries or disorder in open quantum systems},\ }\href
  {https://doi.org/10.1103/PhysRevLett.122.015701} {\bibfield  {journal}
  {\bibinfo  {journal} {Phys. Rev. Lett.}\ }\textbf {\bibinfo {volume} {122}},\
  \bibinfo {pages} {015701} (\bibinfo {year} {2019})}\BibitemShut {NoStop}%
\bibitem [{\citenamefont {Ke\ss{}ler}\ \emph {et~al.}(2021)\citenamefont
  {Ke\ss{}ler}, \citenamefont {Kongkhambut}, \citenamefont {Georges},
  \citenamefont {Mathey}, \citenamefont {Cosme},\ and\ \citenamefont
  {Hemmerich}}]{open_DTC_5}%
  \BibitemOpen
  \bibfield  {author} {\bibinfo {author} {\bibfnamefont {H.}~\bibnamefont
  {Ke\ss{}ler}}, \bibinfo {author} {\bibfnamefont {P.}~\bibnamefont
  {Kongkhambut}}, \bibinfo {author} {\bibfnamefont {C.}~\bibnamefont
  {Georges}}, \bibinfo {author} {\bibfnamefont {L.}~\bibnamefont {Mathey}},
  \bibinfo {author} {\bibfnamefont {J.~G.}\ \bibnamefont {Cosme}},\ and\
  \bibinfo {author} {\bibfnamefont {A.}~\bibnamefont {Hemmerich}},\ }\bibfield
  {title} {\bibinfo {title} {Observation of a dissipative time crystal},\
  }\href {https://doi.org/10.1103/PhysRevLett.127.043602} {\bibfield  {journal}
  {\bibinfo  {journal} {Phys. Rev. Lett.}\ }\textbf {\bibinfo {volume} {127}},\
  \bibinfo {pages} {043602} (\bibinfo {year} {2021})}\BibitemShut {NoStop}%
\bibitem [{\citenamefont {Taheri}\ \emph {et~al.}(2022)\citenamefont {Taheri},
  \citenamefont {Matsko}, \citenamefont {Maleki},\ and\ \citenamefont
  {Sacha}}]{open_DTC_6}%
  \BibitemOpen
  \bibfield  {author} {\bibinfo {author} {\bibfnamefont {H.}~\bibnamefont
  {Taheri}}, \bibinfo {author} {\bibfnamefont {A.~B.}\ \bibnamefont {Matsko}},
  \bibinfo {author} {\bibfnamefont {L.}~\bibnamefont {Maleki}},\ and\ \bibinfo
  {author} {\bibfnamefont {K.}~\bibnamefont {Sacha}},\ }\bibfield  {title}
  {\bibinfo {title} {All-optical dissipative discrete time crystals},\ }\href
  {https://doi.org/10.1038/s41467-022-28462-x} {\bibfield  {journal} {\bibinfo
  {journal} {Nat. Commun.}\ }\textbf {\bibinfo {volume} {13}},\ \bibinfo
  {pages} {848} (\bibinfo {year} {2022})}\BibitemShut {NoStop}%
\bibitem [{\citenamefont {Saha}\ and\ \citenamefont
  {Bhattacharyya}(2024)}]{open_DTC_7}%
  \BibitemOpen
  \bibfield  {author} {\bibinfo {author} {\bibfnamefont {S.}~\bibnamefont
  {Saha}}\ and\ \bibinfo {author} {\bibfnamefont {R.}~\bibnamefont
  {Bhattacharyya}},\ }\bibfield  {title} {\bibinfo {title} {Prethermal discrete
  time crystal in driven-dissipative dipolar systems},\ }\href
  {https://doi.org/10.1103/PhysRevA.109.012208} {\bibfield  {journal} {\bibinfo
   {journal} {Phys. Rev. A}\ }\textbf {\bibinfo {volume} {109}},\ \bibinfo
  {pages} {012208} (\bibinfo {year} {2024})}\BibitemShut {NoStop}%
\bibitem [{\citenamefont {Das}\ \emph {et~al.}(2026)\citenamefont {Das},
  \citenamefont {Saha},\ and\ \citenamefont {Bhattacharyya}}]{Das2026}%
  \BibitemOpen
  \bibfield  {author} {\bibinfo {author} {\bibfnamefont {G.}~\bibnamefont
  {Das}}, \bibinfo {author} {\bibfnamefont {S.}~\bibnamefont {Saha}},\ and\
  \bibinfo {author} {\bibfnamefont {R.}~\bibnamefont {Bhattacharyya}},\
  }\bibfield  {title} {\bibinfo {title} {Environment-assisted discrete time
  crystals in noninteracting quantum systems},\ }\href@noop {} {\bibfield
  {journal} {\bibinfo  {journal} {Phys. Rev. A}\ }\textbf {\bibinfo {volume}
  {113}},\ \bibinfo {pages} {042216} (\bibinfo {year} {2026})}\BibitemShut
  {NoStop}%
\bibitem [{\citenamefont {Manzano~Paule}(2018)}]{manzano2018thermodynamics}%
  \BibitemOpen
  \bibfield  {author} {\bibinfo {author} {\bibfnamefont {G.}~\bibnamefont
  {Manzano~Paule}},\ }\href {https://doi.org/10.1007/978-3-319-93964-3} {\emph
  {\bibinfo {title} {Thermodynamics and Synchronization in Open Quantum
  Systems}}},\ Springer Theses\ (\bibinfo  {publisher} {Springer International
  Publishing},\ \bibinfo {year} {2018})\BibitemShut {NoStop}%
\bibitem [{\citenamefont {Turkeshi}\ \emph {et~al.}(2021)\citenamefont
  {Turkeshi}, \citenamefont {Biella}, \citenamefont {Fazio}, \citenamefont
  {Dalmonte},\ and\ \citenamefont {Schir{\'o}}}]{Turkeshi2021}%
  \BibitemOpen
  \bibfield  {author} {\bibinfo {author} {\bibfnamefont {X.}~\bibnamefont
  {Turkeshi}}, \bibinfo {author} {\bibfnamefont {A.}~\bibnamefont {Biella}},
  \bibinfo {author} {\bibfnamefont {R.}~\bibnamefont {Fazio}}, \bibinfo
  {author} {\bibfnamefont {M.}~\bibnamefont {Dalmonte}},\ and\ \bibinfo
  {author} {\bibfnamefont {M.}~\bibnamefont {Schir{\'o}}},\ }\bibfield  {title}
  {\bibinfo {title} {Measurement-induced entanglement transitions in the
  quantum {Ising} chain: From infinite to zero clicks},\ }\href@noop {}
  {\bibfield  {journal} {\bibinfo  {journal} {Phys. Rev. B}\ }\textbf {\bibinfo
  {volume} {103}},\ \bibinfo {pages} {224210} (\bibinfo {year}
  {2021})}\BibitemShut {NoStop}%
\bibitem [{\citenamefont {Turkeshi}\ \emph {et~al.}(2022)\citenamefont
  {Turkeshi}, \citenamefont {Dalmonte}, \citenamefont {Fazio},\ and\
  \citenamefont {Schir\`o}}]{Ising_2}%
  \BibitemOpen
  \bibfield  {author} {\bibinfo {author} {\bibfnamefont {X.}~\bibnamefont
  {Turkeshi}}, \bibinfo {author} {\bibfnamefont {M.}~\bibnamefont {Dalmonte}},
  \bibinfo {author} {\bibfnamefont {R.}~\bibnamefont {Fazio}},\ and\ \bibinfo
  {author} {\bibfnamefont {M.}~\bibnamefont {Schir\`o}},\ }\bibfield  {title}
  {\bibinfo {title} {Entanglement transitions from stochastic resetting of
  non-{Hermitian} quasiparticles},\ }\href
  {https://doi.org/10.1103/PhysRevB.105.L241114} {\bibfield  {journal}
  {\bibinfo  {journal} {Phys. Rev. B}\ }\textbf {\bibinfo {volume} {105}},\
  \bibinfo {pages} {L241114} (\bibinfo {year} {2022})}\BibitemShut {NoStop}%
\bibitem [{\citenamefont {Le~Gal}\ \emph {et~al.}(2024)\citenamefont {Le~Gal},
  \citenamefont {Turkeshi},\ and\ \citenamefont {Schir\`o}}]{Ising_3}%
  \BibitemOpen
  \bibfield  {author} {\bibinfo {author} {\bibfnamefont {Y.}~\bibnamefont
  {Le~Gal}}, \bibinfo {author} {\bibfnamefont {X.}~\bibnamefont {Turkeshi}},\
  and\ \bibinfo {author} {\bibfnamefont {M.}~\bibnamefont {Schir\`o}},\
  }\bibfield  {title} {\bibinfo {title} {Entanglement dynamics in monitored
  systems and the role of quantum jumps},\ }\href
  {https://doi.org/10.1103/PRXQuantum.5.030329} {\bibfield  {journal} {\bibinfo
   {journal} {PRX Quantum}\ }\textbf {\bibinfo {volume} {5}},\ \bibinfo {pages}
  {030329} (\bibinfo {year} {2024})}\BibitemShut {NoStop}%
\bibitem [{\citenamefont {Piccitto}\ \emph {et~al.}(2022)\citenamefont
  {Piccitto}, \citenamefont {Russomanno},\ and\ \citenamefont
  {Rossini}}]{diff_unravelling_Ising}%
  \BibitemOpen
  \bibfield  {author} {\bibinfo {author} {\bibfnamefont {G.}~\bibnamefont
  {Piccitto}}, \bibinfo {author} {\bibfnamefont {A.}~\bibnamefont
  {Russomanno}},\ and\ \bibinfo {author} {\bibfnamefont {D.}~\bibnamefont
  {Rossini}},\ }\bibfield  {title} {\bibinfo {title} {Entanglement transitions
  in the quantum {Ising} chain: A comparison between different unravelings of
  the same {Lindbladian}},\ }\href
  {https://doi.org/10.1103/PhysRevB.105.064305} {\bibfield  {journal} {\bibinfo
   {journal} {Phys. Rev. B}\ }\textbf {\bibinfo {volume} {105}},\ \bibinfo
  {pages} {064305} (\bibinfo {year} {2022})}\BibitemShut {NoStop}%
\bibitem [{\citenamefont {M{\"u}ller}\ \emph {et~al.}(2025)\citenamefont
  {M{\"u}ller}, \citenamefont {Buchhold},\ and\ \citenamefont
  {Diehl}}]{muller2025monitored}%
  \BibitemOpen
  \bibfield  {author} {\bibinfo {author} {\bibfnamefont {T.}~\bibnamefont
  {M{\"u}ller}}, \bibinfo {author} {\bibfnamefont {M.}~\bibnamefont
  {Buchhold}},\ and\ \bibinfo {author} {\bibfnamefont {S.}~\bibnamefont
  {Diehl}},\ }\bibfield  {title} {\bibinfo {title} {Monitored interacting
  {Dirac} fermions},\ }\href@noop {} {\bibfield  {journal} {\bibinfo  {journal}
  {Phys. Rev. B}\ }\textbf {\bibinfo {volume} {111}},\ \bibinfo {pages}
  {174201} (\bibinfo {year} {2025})}\BibitemShut {NoStop}%
\bibitem [{\citenamefont {Starchl}\ \emph {et~al.}(2025)\citenamefont
  {Starchl}, \citenamefont {Fischer},\ and\ \citenamefont
  {Sieberer}}]{starchl2025generalized}%
  \BibitemOpen
  \bibfield  {author} {\bibinfo {author} {\bibfnamefont {E.}~\bibnamefont
  {Starchl}}, \bibinfo {author} {\bibfnamefont {M.~H.}\ \bibnamefont
  {Fischer}},\ and\ \bibinfo {author} {\bibfnamefont {L.~M.}\ \bibnamefont
  {Sieberer}},\ }\bibfield  {title} {\bibinfo {title} {Generalized {Zeno}
  effect and entanglement dynamics induced by fermion counting},\ }\href@noop
  {} {\bibfield  {journal} {\bibinfo  {journal} {PRX Quantum}\ }\textbf
  {\bibinfo {volume} {6}},\ \bibinfo {pages} {030302} (\bibinfo {year}
  {2025})}\BibitemShut {NoStop}%
\bibitem [{\citenamefont {Chatterjee}\ and\ \citenamefont
  {Modak}(2025)}]{Chatterjee2025}%
  \BibitemOpen
  \bibfield  {author} {\bibinfo {author} {\bibfnamefont {P.}~\bibnamefont
  {Chatterjee}}\ and\ \bibinfo {author} {\bibfnamefont {R.}~\bibnamefont
  {Modak}},\ }\bibfield  {title} {\bibinfo {title} {Measurement-induced phase
  transition in periodically driven free-fermionic systems},\ }\href@noop {}
  {\bibfield  {journal} {\bibinfo  {journal} {Phys. Rev. B}\ }\textbf {\bibinfo
  {volume} {112}},\ \bibinfo {pages} {024304} (\bibinfo {year}
  {2025})}\BibitemShut {NoStop}%
\bibitem [{\citenamefont {Matsubara}\ \emph {et~al.}(2025)\citenamefont
  {Matsubara}, \citenamefont {Yamamoto},\ and\ \citenamefont
  {Koga}}]{quasiperiodic_measurement}%
  \BibitemOpen
  \bibfield  {author} {\bibinfo {author} {\bibfnamefont {T.}~\bibnamefont
  {Matsubara}}, \bibinfo {author} {\bibfnamefont {K.}~\bibnamefont
  {Yamamoto}},\ and\ \bibinfo {author} {\bibfnamefont {A.}~\bibnamefont
  {Koga}},\ }\bibfield  {title} {\bibinfo {title} {Measurement-induced phase
  transitions for free fermions in a quasiperiodic potential},\ }\href
  {https://doi.org/10.1103/3zfd-3hqt} {\bibfield  {journal} {\bibinfo
  {journal} {Phys. Rev. B}\ }\textbf {\bibinfo {volume} {112}},\ \bibinfo
  {pages} {054309} (\bibinfo {year} {2025})}\BibitemShut {NoStop}%
\bibitem [{\citenamefont {Delmonte}\ \emph {et~al.}(2026)\citenamefont
  {Delmonte}, \citenamefont {Li}, \citenamefont {Fazio},\ and\ \citenamefont
  {Romito}}]{most_prob_trj_2026}%
  \BibitemOpen
  \bibfield  {author} {\bibinfo {author} {\bibfnamefont {A.}~\bibnamefont
  {Delmonte}}, \bibinfo {author} {\bibfnamefont {Z.}~\bibnamefont {Li}},
  \bibinfo {author} {\bibfnamefont {R.}~\bibnamefont {Fazio}},\ and\ \bibinfo
  {author} {\bibfnamefont {A.}~\bibnamefont {Romito}},\ }\bibfield  {title}
  {\bibinfo {title} {{Measurement-induced phase transition in interacting
  bosons from most likely quantum trajectory}},\ }\href
  {https://doi.org/10.21468/SciPostPhys.20.4.109} {\bibfield  {journal}
  {\bibinfo  {journal} {SciPost Phys.}\ }\textbf {\bibinfo {volume} {20}},\
  \bibinfo {pages} {109} (\bibinfo {year} {2026})}\BibitemShut {NoStop}%
\bibitem [{\citenamefont {Leung}\ \emph {et~al.}(2025)\citenamefont {Leung},
  \citenamefont {Meidan},\ and\ \citenamefont
  {Romito}}]{partial_post_selection}%
  \BibitemOpen
  \bibfield  {author} {\bibinfo {author} {\bibfnamefont {C.~Y.}\ \bibnamefont
  {Leung}}, \bibinfo {author} {\bibfnamefont {D.}~\bibnamefont {Meidan}},\ and\
  \bibinfo {author} {\bibfnamefont {A.}~\bibnamefont {Romito}},\ }\bibfield
  {title} {\bibinfo {title} {Theory of free fermions dynamics under partial
  postselected monitoring},\ }\href
  {https://doi.org/10.1103/PhysRevX.15.021020} {\bibfield  {journal} {\bibinfo
  {journal} {Phys. Rev. X}\ }\textbf {\bibinfo {volume} {15}},\ \bibinfo
  {pages} {021020} (\bibinfo {year} {2025})}\BibitemShut {NoStop}%
\bibitem [{\citenamefont {Caro}\ and\ \citenamefont
  {G\'amez}(2026)}]{DTC_projective}%
  \BibitemOpen
  \bibfield  {author} {\bibinfo {author} {\bibfnamefont {C.}~\bibnamefont
  {Caro}}\ and\ \bibinfo {author} {\bibfnamefont {F.}~\bibnamefont {G\'amez}},\
  }\bibfield  {title} {\bibinfo {title} {Intermittent time-crystalline dynamics
  under stochastic measurements},\ }\href {https://doi.org/10.1103/y9yn-x6x2}
  {\bibfield  {journal} {\bibinfo  {journal} {Phys. Rev. A}\ }\textbf {\bibinfo
  {volume} {114}},\ \bibinfo {pages} {032422} (\bibinfo {year}
  {2026})}\BibitemShut {NoStop}%
\bibitem [{\citenamefont {Jacobs}\ and\ \citenamefont
  {Steck}(2006)}]{measurement_def}%
  \BibitemOpen
  \bibfield  {author} {\bibinfo {author} {\bibfnamefont {K.}~\bibnamefont
  {Jacobs}}\ and\ \bibinfo {author} {\bibfnamefont {D.~A.}\ \bibnamefont
  {Steck}},\ }\bibfield  {title} {\bibinfo {title} {A straightforward
  introduction to continuous quantum measurement},\ }\href
  {https://doi.org/10.1080/00107510601101934} {\bibfield  {journal} {\bibinfo
  {journal} {Contemp. Phys.}\ }\textbf {\bibinfo {volume} {47}},\ \bibinfo
  {pages} {279} (\bibinfo {year} {2006})}\BibitemShut {NoStop}%
\bibitem [{\citenamefont {Cao}\ \emph {et~al.}(2019)\citenamefont {Cao},
  \citenamefont {Tilloy},\ and\ \citenamefont {De~Luca}}]{cao2019entanglement}%
  \BibitemOpen
  \bibfield  {author} {\bibinfo {author} {\bibfnamefont {X.}~\bibnamefont
  {Cao}}, \bibinfo {author} {\bibfnamefont {A.}~\bibnamefont {Tilloy}},\ and\
  \bibinfo {author} {\bibfnamefont {A.}~\bibnamefont {De~Luca}},\ }\bibfield
  {title} {\bibinfo {title} {Entanglement in a fermion chain under continuous
  monitoring},\ }\href@noop {} {\bibfield  {journal} {\bibinfo  {journal}
  {SciPost Phys.}\ }\textbf {\bibinfo {volume} {7}},\ \bibinfo {pages} {024}
  (\bibinfo {year} {2019})}\BibitemShut {NoStop}%
\bibitem [{\citenamefont {Yoshida}(2024)}]{steady_state_lindblad}%
  \BibitemOpen
  \bibfield  {author} {\bibinfo {author} {\bibfnamefont {H.}~\bibnamefont
  {Yoshida}},\ }\bibfield  {title} {\bibinfo {title} {Uniqueness of steady
  states of {Gorini-Kossakowski-Sudarshan-Lindblad} equations: A simple
  proof},\ }\href {https://doi.org/10.1103/PhysRevA.109.022218} {\bibfield
  {journal} {\bibinfo  {journal} {Phys. Rev. A}\ }\textbf {\bibinfo {volume}
  {109}},\ \bibinfo {pages} {022218} (\bibinfo {year} {2024})}\BibitemShut
  {NoStop}%
\end{thebibliography}%

\end{document}